\documentclass[prb,aps,showpacs,twocolumn,preprintnumbers,amsmath,amssymb,superscriptaddress,nofootinbib]{revtex4-1}
\usepackage{amsmath,amssymb,amsfonts}
\usepackage{graphicx}
\usepackage[colorlinks=True,linkcolor=black,citecolor=blue,urlcolor=blue]{hyperref}
\usepackage{xcolor}
\usepackage{chngcntr}
\usepackage{braket}
\usepackage{hyperref}
\usepackage{nicefrac}
\usepackage{bm}
\usepackage{float}
\usepackage{ulem}
\usepackage{multirow}
\usepackage{booktabs}

\newcommand{\bs}[1]{\boldsymbol{#1}}

\begin{document}

\title{Transport signatures of valley polarization in graphene multilayers: In-plane linear magnetoconductivity vs anomalous Hall effect}

\author{Fernando Peñaranda}
\affiliation{Donostia International Physics Center, P. Manuel de Lardizabal 4, 20018 Donostia-San Sebastian, Spain}
\author{Fernando de Juan}
\affiliation{Donostia International Physics Center, P. Manuel de Lardizabal 4, 20018 Donostia-San Sebastian, Spain}

\affiliation{IKERBASQUE, Basque Foundation for Science, Maria Diaz de Haro 3, 48013 Bilbao, Spain}

\date{\today}

\begin{abstract}
In two-dimensional materials where interacting Fermi pockets occur in valleys related by time-reversal symmetry, a spontaneous valley imbalance results in a novel state known as an orbital magnet. Due to the breaking of time-reversal symmetry, this state can be probed in transport experiments by the violation of Onsager relations, most often done through the anomalous Hall effect (AHE). Here we propose that odd-in-field, in-plane linear magnetoconductivity (LMC) is an alternative probe of valley polarization which can occur even when the AHE vanishes. In multilayer structures, the effect originates from in-plane orbital moments and Berry curvatures enabled by interlayer tunneling and dominates over the spin response. After a classification of many recently studied multilayers, we focus on two valley polarized examples: twisted bilayer graphene, where LMC is finite but the AHE vanishes unless additional symmetry breaking from the substrate is present, and rhombohedral graphene multilayers, where LMC and AHE both track valley polarization because they have the same symmetry. Using self-consistent Hartree-Fock and semiclassical transport calculations, we present detailed predictions of LMR for these two examples and analyze the implications for recent experiments.   
\end{abstract}

\maketitle
\begin{figure}[t]
    \centering
    \includegraphics[width=1\linewidth]{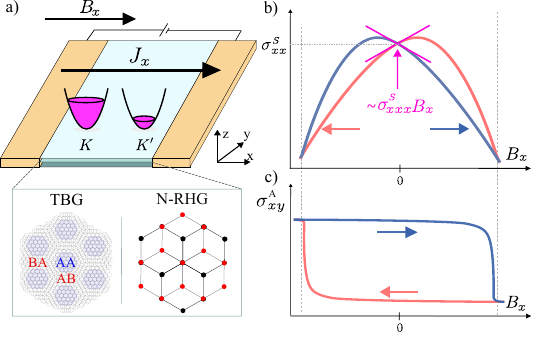}
    \caption{(a) Sketch of the proposed two-contacts measurement setup for the observation of in-plane LMC. In a valley-polarized state, where the Fermi-surface pockets exhibit unequal fillings (depicted in pink), an in-plane magnetic field $B_x$ results in a time-odd $\sigma^{S}_{xxx}$ correction to $\sigma^{S}_{xx}$. The inset depicts two representative systems studied in this work:  TBG and N-layer RG belonging to class II and III in Table~\ref{tab:1}, respectively, whose symmetries in the presence of VP enable a finite in-plane response.  (b-c) Schematic showing the two symmetry-nonequivalent dominant hysteresis effects that serve as transport probes of TRS breaking due to (b) the in-plane LMC and (c) the standard AHE, where colors represents distinct magnetic-field sweep directions through the hysteresis loop. A perturbative expansion around $|\vec B| = 0$ yields $\sigma^{S}_{xxx}$ as the leading perturbation to the $B=0$ linear conductivity $\sigma^{S}_{xx}$ (depicted in pink). }
    \label{fig:sketch}
\end{figure}

\emph{Introduction} - The phenomenon of magnetism is generally pictured as the ordering of spin magnetic moments, of either itinerant or localized nature~\cite{Blundell01}. In metallic systems where disjoint Fermi surfaces, known as valleys, occur in pairs related by time-reversal symmetry (TRS), an altogether different type of magnetism is possible, driven by a spontaneous population imbalance between valleys \cite{Braz18,Liu:PRX19,Bultinck:PRL20}. These valley polarized states, where magnetism is predominantly of orbital origin, have been recently realized in a number of two-dimensional material heterostructures   with trigonal or hexagonal symmetry, where Fermi pockets occur around the $K$ and $K'=-K$ valleys (see Fig. \ref{fig:sketch}a). These systems, which include twisted bilayer graphene~\cite{Sharpe19,Serlin20,Lu19,Polshyn20}, twisted transition metal dichalcogenides~\cite{Anderson23,Cai23,Zeng23}, and more recently rhombohedral graphene~\cite{Zhou:21N, Zhou:21N2,Liu:NN24, Sha:S24, Choi:N25,Han:N23, Han:NN24,Qin:A26, Deng:A25, Zheng:PRL25,Zhou:NC24}, all feature flat bands near the Fermi level where interactions are enhanced. These lead to variety of correlated phases, among which such valley polarized states are frequently
found. 

Transport experiments are a powerful probe to demonstrate the existence of these valley polarized phases and map them across phase diagrams. The breaking of TRS can be detected in transport using the Onsager reciprocity relation \cite{Shtrikman65}
\begin{align}
\sigma_{ij}(\vec B,\phi) = \sigma_{ji}(-\vec B,-\phi)    
\end{align}
where $\phi$ represents any order parameter which is odd under TRS. This relation allows to separate the time-even and odd components of the conductivity $\sigma_{ij} = \sigma_{ij}^{\rm even} +\sigma_{ij}^{\rm odd}$ as 

\begin{align}
    \sigma_{ij}^{\rm even} &= \sigma_{ij}^{S} + \sigma_{ijk}^{A}B_k+\ldots \\
    \sigma_{ij}^{\rm odd} &= \sigma_{ij}^{A} + \sigma_{ijk}^{S}B_k+\ldots
\end{align}
where $S/A$ denotes symmetric or antisymmetric with respect to $i \leftrightarrow j$. $\sigma_{ij}^{\rm even}$, composed of Drude ($\sigma_{ij}^S$) and Hall ($\sigma_{ijk}^A$) conductivities, is invariant under TRS, while $\sigma_{ij}^{\rm odd}$ flips sign and hence should vanish. The breaking of TRS enables the observation of $\sigma_{ij}^{\rm odd}$, not only through an anomalous Hall effect $\sigma_{ij}^{A}$, but also through longitudinal transport that is linear in magnetic field around $\vec B=0$, described by the linear magnetoconductivity (LMC) $\sigma_{ijk}^{S}$. However, while $\sigma_{ij}^{A}$ has the symmetry of a magnetic moment and is allowed only in ferromagnetic states, the LMC tensor $\sigma_{ijk}^{S}$~\cite{Shtrikman65,Cortijo16,Zyuzin21,Sunko:PRB25,Das25} is allowed in all piezomagnetic crystals \cite{Dzialoshinskii58}, regardless of whether they have a ferromagnetic moment, and hence probes the breaking of time-reversal symmetry in a broader class of systems.    

In this work, we analyze how linear magnetoconductivity probes valley polarization in a wide range of experimentally realized 2D heterostructures. A general symmetry analysis shows there are three classes of structures: those that show neither LMC nor AHE, those with only in-plane LMC, and those with both LMC and AHE. Notably, in-plane LMC in such multilayer structures originates from in-plane orbital moments~\cite{Drigo20,Murakami20,Ghorai25} and Berry curvatures~\cite{Kim21} due to interlayer coupling and dominates over the Zeeman effect. We have computed in-plane LMC for a representative case in the two non-trivial classes. In twisted bilayer graphene (TBG), where AHE vanishes due to an in-plane two-fold rotation axis $C_{2x}$, $\sigma_{xxx}^{S}$ is the only probe of valley polarization. In the presence of a substrate, the AHE becomes activated but its sign depends on substrate details and may average out in transport, while the LMC does not. In rhombohedral graphene, where LMC has been observed \cite{Lee22}, both AHE and LMC probe valley polarization, although they have in general different dependencies on filling. Our detailed analysis shows, however, that these effects only occur in the quarter metal states, contrary to the theoretical interpretation in Ref. \cite{Lee22}. We substantiate all our claims by detailed symmetry analysis and explicit semiclassical calculations in Hartree-Fock mean field ground states for effective models for TBG and rhombohedral graphene.

\section{Odd-in-field linear magnetoconductivity \label{OMRsym}}

The LMC tensor $\sigma_{ijk}^{S}$ is defined as the contribution to the conductivity which is linear in $\vec B$ around $\vec B =0$, and should not be confused with the high field scaling $\sigma_{ij}\propto|\vec B|$ present in some non-magnetic systems \cite{Abrikosov00}. LMC can only be non-zero when time-reversal symmetry is broken due to a time-reversal odd order parameter $\phi$. While $\phi (\vec B)$ may depend on the applied $\vec B$ in general, this dependence only corrects the conductivity to higher powers in $\vec B$, so that $\sigma_{ijk}^{S}$ can be understood as a response in the presence of a static, frozen $\phi$. This effect is thus different from anisotropic magnetoresistance (AMR) \cite{Ritzinger23} in ferromagnets, which measures the conductivity dependence on magnetization when it is saturated by the applied field. LMC is in the opposite limit near zero $\vec B$ when the magnetic order is independent of magnetization. Also note in this work we consider bulk LMC. Edge transport in magnetic fields can lead to odd-in-B resistance which is opposite in the two sides of the sample. This signal is picked up in four terminal measurements that measure a single side, but averages to zero when signals from both sides are added, or in a two terminal measurement~\cite{Takiguchi22,Sahani24}. Such edge LMC is not forbidden by Onsager relations and can occur in non-magnetic materials~\cite{Wang25}. 

Depending on the symmetry of $\phi$, $\vec B$ may or may not couple to $\phi$. If it does, at sufficiently large values of $\vec B$ the sign of $\phi$ can be flipped. In this case, as $\vec B$ is swept in experiments a hysteresis loop will be observed with a characteristic crossing at $\vec B=0$ for the two branches corresponding to the two signs of $\phi$, since $\sigma_{ijk}^{S}(-\phi) = -\sigma_{ijk}^{S}(\phi)$ (see Fig. \ref{fig:sketch}b). LMC is indeed observed as such crossing in many magnetic systems \cite{Fujita15,Wang20,Feng22,Jiang21,Lazrak25} and because of this it is also called antisymmetric or butterfly magnetoresistance. Note this type of hysteresis loop is very different from the AHE (Fig. \ref{fig:sketch}c). For more complex magnetic orders it might be the case that symmetry forbids the coupling to $\phi$, in which case only a single slope will be observed with no hysteresis. 

The allowed components of the LMC tensor are constrained by the magnetic point group (see for example MTENSOR in the Bilbao Crystallographic Server\cite{MTENSOR}) and can be non-zero only in piezomagnetic groups. An intuitive way to understand these constraints is that TRS combined with a point group element enforces extra Onsager-type constraints in the conductivity tensor. If $S \mathcal{T}$, where $S$ is an order 2 point group symmetry, leaves $\phi$ invariant, then $S \phi = -\phi$ and the Onsager constraint is
\begin{align}
S \sigma(\bs B, \phi) S^\dagger = \sigma^T(- S \bs B,    \phi), \label{generalizedOnsager} 
\end{align}
which implies the components of $\sigma_{ijk}^{S}$ which are invariant under $S$ must vanish. For example, if $\phi$ preserves $\mathcal{PT}$ symmetry, then all components vanish because $J_iE_jB_k$ is invariant under $P$. Similarly, if $S=C_{2z}$, then $\sigma_{ijk}^{S}=0$ when all indices are in-plane.

\subsection{Symmetry analysis of two-dimensional heterostructures}\label{SymAnalysis}

\begin{table*}[]
    \centering
    \setlength{\tabcolsep}{5.6pt}
    \begin{tabular}{|c|c|c|c|c|c|c|c|c|}\hline
     Class & Point group & VP IR & Magnetic PG & Examples & $\sigma^{S}_{xxx}$ & $\sigma^{S}_{yyy}$ & $\sigma^{S}_{xxz}$ & $\sigma_{xy}^A$ \\ \hline \hline
     
     \multirow{3}{*}{I} & $D_{6h}$ ($\rm 6/mmm$) & $B_{1u}$ & $\rm 6'/mmm'$ ($C_{2x}\mathcal{T}$,$C_{2z}\mathcal{T}$,$I\mathcal{T}$,$M_{y}\mathcal{T}$) &  1L Graphene & 0 & 0 & 0  & 0
     \\  \cline{2-9}
     
     & $D_{6h}$ ($\rm 6/mmm$) & $B_{2u}$ & $\rm 6'/mm'm$ ($C_{2y}\mathcal{T}$,$C_{2z}\mathcal{T}$,$I\mathcal{T}$,$M_{x}\mathcal{T}$) &  TTG & 0 & 0 & 0  & 0
     \\   \cline{2-9}
     
     &  $D_{3d}$ ($\rm \bar{3}m$)& $A_{1u}$ & $\rm \bar{3}'m'$ ($I\mathcal{T}$,$M_{y}\mathcal{T}$) & 2N Bernal, 2N+1 RG & 0 & 0 & 0  &0
     \\ \hline \hline
    \multirow{5}{*}{II} & $D_{6}$ ($\rm 622$) & $B_1$ & $\rm 6'22'$ ($C_{2y}\mathcal{T}$,$C_{2z}\mathcal{T}$) & TBG & \checkmark & 0 & 0  &0
     \\   \cline{2-9}
    & $D_{3}$ ($\rm 32$) & $A_1$ & $\rm 32$ & AB-AB TDBG & \checkmark & 0 & 0 &0   
   \\ \cline{2-9}
     &  $D_{3h}$ ($\rm \bar{6}m2$) & $A''_1$ & $\rm \bar{6}'m'2$ ($M_z\mathcal{T}$,$M_{y}\mathcal{T}$) & AB tWSe2 & \checkmark & 0 & 0 & 0\\ \cline{2-9}
     & $C_{6v}$ ($\rm 6mm$) & $B_2$ & $\rm 6'm'm$ ($C_{2z}\mathcal{T}$,$M_x \mathcal{T}$) & TTG in $E_z$ & 0 & \checkmark & 0  &0 
     \\ \cline{2-9}
     & $C_{6}$ ($\rm 6$) & $B$ & $\rm 6'$ ($C_{2z}\mathcal{T}$) & TBG in $E_z$ & \checkmark & \checkmark & 0  &0 
     \\ \hline \hline
        
     \multirow{5}{*}{III} & $D_{3h}$ ($\rm \bar{6}m2$) & $A'_2$ & $\rm \bar{6}m'2'$ ($C_{2x}\mathcal{T}$,$M_{y}\mathcal{T}$) & 1L TMD, 2N+1 Bernal & 0 & 0 & \checkmark  &\checkmark
     \\ \cline{2-9}
    
    & $D_{3}$ ($\rm 32$) & $A_2$ & $\rm32'$ ($C_{2y}\mathcal{T}$) & AB-BA TDBG & \checkmark & 0 & \checkmark  &\checkmark   
     \\ \cline{2-9}
     & $D_{3d}$ ($\rm \bar{3}m$)& $A_{2g}$ &$\rm \bar{3}m'$ ($C_{2y}\mathcal{T}$,$M_{y}\mathcal{T}$) & AA tWSe$_2$ & \checkmark & 0 & \checkmark  &\checkmark
     \\ \cline{2-9}
     & $C_{3v}$ ($\rm 3m$) & $A_2$ & $\rm 3m'$ ($M_{y}\mathcal{T}$) & 2N RG & \checkmark & 0 & \checkmark  &\checkmark \\ \cline{2-9}
     & $C_{3}$ ($\rm 3$) & $A_1$ & $\rm 3$  & TDBG in $E_z$ & \checkmark & \checkmark & \checkmark  &\checkmark \\ \hline
    \end{tabular}
    \caption{Symmetry of 2D heterostructures with valley polarization (VP). In all systems we choose the $x$ along the $\Gamma-M$ direction. VP IR denotes the irreducible representation under which VP transforms. Intervalley symmetries which combined with time-reversal remain a symmetry in the valley polarized state are shown in parenthesis. Note in tWSe$_2$ some symmetries are approximate, see text for details. Entries are grouped according to whether neither LMC or AHE are allowed (class I), only LMC is allowed (class II) or both are allowed (class III). In the presence of electric field $E_z$, all examples become $3m'$($M_y\mathcal{T}$) except TTG ($6'm'm$), TBG ($6'$ ) and TDBG ($3$).}
    \label{tab:1}
\end{table*}

In this work, we consider $\sigma_{ijk}^{S}$ as a probe of valley polarization in two-dimensional heterostructures. Here we present the symmetry anaylsis of $\sigma_{ijk}^{S}$ for the most common 2D heterostructures including twisted graphene multilayers, twisted transition metal dichalcogenides and rhombohedral graphene multilayers. Note that we do not make any assumptions on the origin of valley polarization or discuss under what conditions it is stable, but rather just analyze its symmetry and effects on transport. For ease of comparison, we take a coordinate system where $\Gamma-M$ is along the $\hat x$ direction, and hence the lattice vectors (or moire lattice vectors when relevant) are aligned with $\hat y$ and $C_3$ related directions. We perform the symmetry analysis assuming no spin-orbit coupling so that the spin degree of freedom does not transform under the point group. By identifying the point group operations in each system which interchange the valleys, and hence are broken by valley polarization, we can determine under which time-odd irreducible representation (irrep) such valley polarization transforms. This then determines the magnetic point group and the allowed components of LMC and AHE. Since valley polarization preserves $C_3$ in all cases, and we consider $J_i$ and $E_i$ only in the plane, there are only three independent components of LMC, $\sigma^{S}_{xxx}=-\sigma^{S}_{yyx}= -\sigma_{xyy}^{S}$, $\sigma^{S}_{yyy}=-\sigma^{S}_{xxy}= -\sigma_{yxx}^{S}$ and $\sigma^{S}_{xxz}=\sigma^{S}_{yyz}$. For simplicity these are just denoted as  $\sigma^{S}_{xxx}$, $\sigma^{S}_{yyy}$ and $\sigma^{S}_{xxz}$ in the rest of this work. For AHE there is only $\sigma_{xy}^A$. In the presence of $C_3$ symmetry, LMC when the magnetic field is applied out of the plane $\sigma^{S}_{xxz}=\sigma^{S}_{yyz}$ transforms in the same way as $\sigma_{xy}^A$ \cite{Das25}, so either they are both allowed or forbidden. However, LMC when the field is in the plane has a different symmetry and can be allowed even when $\sigma_{xy}^A=0$. This gives rise to three classes of systems, depending on whether neither LMC nor AHE is allowed, only in-plane LMC is allowed, or both AHE and LMC are allowed. The results of our anaylsis are summarized in Table \ref{tab:1}. 

First we consider non-twisted graphene multilayers~\cite{Guinea06,Koshino10}. A single layer of graphene has $D_{6h}$ symmetry and $B_{1u}$ VP, and due to $I\mathcal{T}$ has neither LMC nor AHE. Graphene $N$-layer structures without twist can be stacked in two different configurations~\cite{Malard09}: ABA (Bernal stacking) and ABC (rhombohedral stacking). $2N$-Bernal and $(2N+1)$-rhombohedral graphene (RG) have $D_{3d}$ symmetry and $A_{1u}$ VP, which again shows neither LMC nor AHE due to $I\mathcal{T}$. $(2N+1)$-Bernal graphene has group $D_{3h}$ and VP $A_2'$, which allows only out of plane LMC and AHE. The in-plane LMC is forbidden by the horizontal mirror $M_z$. $2N$-RG has group $C_{3v}$ and VP $A_2$, which does allow both in-plane and out of plane LMC and AHE.

Next we consider twisted graphene multilayers. Twisted bilayer graphene has $D_6$ symmetry~\cite{Zou18} and $B_1$ VP, and shows only in-plane LMC. Symmetric twisted trilayer graphene \cite{Khalaf19,Calugaru21} has group $D_{6h}$ and $B_{2u}$ VP, which has $I\mathcal{T}$ symmetry and vanishing LMC and AHE. Twisted double bilayer graphene \cite{Koshino19} provides a more complex example. Since the bilayer constituents do not have $C_{2z}$ symmetry, there are two possible twisted structures related by a $C_{2z}$ rotation of one layer with respect to the other, AB-AB and AB-BA stackings. In the AB-AB case the point group is $D_3$ and VP is $A_1$, and only in-plane LMC is allowed. For AB-BA, on the other hand the irrep is $A_2$ and both LMC and AHE are allowed. 

Twisted transition metal dichalcogenides such as WSe$_2$ also provide a more complex example. A monolayer TMD has point group $D_{3h}$ and VP $A'_2$ like odd-layer RG. In TMD heterostructures, there are two mechanisms that generate a moire pattern, a relative twist between layers (which may be the same or different TMDs) or lattice mismatch (which only occurs in heterobilayers) \cite{Ruiz-Tijerina19}. In addition, due to the absence of $C_{2z}$ there are also two types of stacking known as AA and AB stacking. The AA twisted homobilayer has exact point group $D_3$ with an intervalley two-fold axis, but the continuum model also has an approximate intravalley inversion symmetry, making the group $D_{3d}$. The AA zero-twist heterobilayer, on the other hand, has exact group $C_{3v}$ with an intervalley mirror, and also an approximate inversion symmetry, making the group again $D_{3d}$. Therefore, at the level of the continuum model both cases can be treated in the same way. The VP irrep is $A_{2g}$ and both LMC and AHE are allowed. The AB stacked case has a similar analysis where the intravalley inversion is replaced by intervalley horizontal mirror $M_z$. The AB twisted homobilayer has exact group $D_{3}$ with intravalley two-fold axis and an approximate group $D_{3h}$, while the AB zero-twist heterobilayer has exact group $C_{3v}$ with intervalley mirror and approximate $D_{3h}$ group. In both cases the the VP irrep is $A''_1$, and only in-plane LMC is allowed. While in this work we will not focus on TMD structures and provide this classification for completeness, it is worth noting that the AHE \cite{Anderson23,Cai23,Zeng23} is seen along $\sigma_{xxz}$\cite{Li21,Xu23} in valley polarized states in these systems. 

Finally, we consider all of these heterostructures in the presence of an applied electric field $E_z$, which is often relevant in experiments where a gate potential is applied. The electric field breaks inversion and all in-plane two-fold axes. In most cases, this leads to the group $C_{3v}$ with both LMC and AHE. An interesting case occurs for the twisted cases TTG and TBG, where the applied field induces in-plane LMC but no AHE. In TDBG the applied field removes the distinction between AB-AB and AB-BA stackings and both LMC and AHE are allowed.
 
To summarize, symmetry analysis reveals there are three different classes of valley polarized heterostructures. In class I, the presence of $PT$ symmetry forbids both the AHE and LMC. In class II, the AHE is forbidden and only in-plane LMC serves as a probe of valley polarization. In the class III, all effects (AHE and in-plane and out of plane LMC) are allowed. This analysis then motivates our study of in-plane LMC as a robust probe of valley polarization. In the next section, we show how in-plane LMC can be computed in a two-dimensional system with a finite extent in the $z$ direction.

\subsection{Semiclassical derivation of LMC}

Within the relaxation time approximation, the Boltzmann formalism\cite{Sundaram99, Xiao10} yields for this transport effect the following semiclassical expression in the canonical ensemble\cite{Shtrikman65, Morimoto16, Zyuzin21, Mandal22, Sunko:PRB25, Das25}
\begin{align}
    \sigma^{S}_{ijk} = \tilde \sigma^{S}_{ijk} + \delta\sigma_{ijk}^{S}, \label{fullsemiclassical}
\end{align}
expressed as the sum of its grand-canonical analogue 
\begin{align}
\tilde \sigma^{S}_{ijk} =&\frac{\tau e^2}{\hbar}  \int_{n \bs k}  f'_n \Big[ \frac{1}{2} \left(v^i_{n} \partial_j + v^j_{n} \partial_i \right)m^k_{n} \nonumber\\
&- v^{ij}_{n} m^k_{n} + e \Big( v^i_{n}  v^j_{n}  \Omega^k_{n} \nonumber \\ 
&- \left( \delta_{jk} v^i_{n} + \delta_{ik} v^j_{n}  \right)\sum_{q \in{x,y}}  v^q_{n} \Omega^q_{n}\Big) \Big], \label{semiclassicalmain}
\end{align}
where the chemical potential $\mu$, rather than the electron density $n$, is held fixed, and a correction term 
\begin{align}
\delta \sigma^{S}_{ijk} = \frac{-e^2 \tau}{N_0}  \int_{n \bs k} \big[\frac{e}{\hbar}\Omega^k_{n} f_n  - m^k_{n} f'_n\big]  \int_{n\bs k}f_n''v^i_{n} v^j_{n}, \label{correctionmain}
\end{align}
where the $\bs k$-dependence has been omitted and summation over repeated indices is implied. The canonical Eq.~\eqref{fullsemiclassical} is chosen for our calculations since it is the one relevant in the experiment. Above, $\tau$ is the relaxation time, $\int_{n \bs k} \equiv \sum_n \int_{\bs k} \frac{d^d k}{(2\pi)^d}$, $\{i,\ j,\ k\} \in \{x,\ y,\ z\}$ denote the cartesian indices, $f_n' = \partial_\varepsilon f_n(\epsilon_{n \bs k}^0)$ is the energy-derivative of the distribution function of band $n$ of the unperturbed ($B = 0$) problem, $N_0 =-\int_{n\bs k} f_n'$, $\hbar v^i_{n} = \partial_i \varepsilon^0_n(\bs k)$, $v^{ij}_{n} = \partial_i v^j_{n} $, $\bs \Omega_{n \bs k } = i \langle \bs \nabla_{\bs k} u_{n \bs k}|\times \bs \nabla_{\bs k} u_{n \bs k } \rangle$ is the Berry curvature with $|u_{n\bs k}\rangle$ denoting the eigenvectors of the crystal Hamiltonian $H|u_{n\bs k}\rangle = \varepsilon^0_{n \bs k} |u_{n \bs k}\rangle$ at $B = 0$,
\begin{equation}
    \bs m_{n \bs k} = -i \frac{e}{2\hbar} \langle \bs \nabla_{\bs k} u_{n \bs k}|\times (H_{\bs k} - \varepsilon^0_{n \bs k}) \bs |\nabla_{\bs k} u_{n \bs k}\rangle -g \frac{\mu_B}{\hbar
    } \langle \bs s\rangle , \label{OMMm}
\end{equation}
denotes the magnetic moment of band $n$, separated into the orbital (OMM) and spin contribution (SMM) with $\bs s =  [s_x,s_y,s_z]$ and $s_i$ the Pauli matrices acting on spin space, $\mu_B$ the Bohr magneton, and $g\approx2$. The full derivation of Eqs. \eqref{semiclassicalmain} and \eqref{correctionmain} is provided in Appendix \ref{AA}. 

\subsection{In-plane LMC in quasi-two dimensions}\label{LGFormalism}
The analysis presented in Sec.~\ref{SymAnalysis} (see Table \ref{tab:1}) concluded that among the three independent tensor components of the LMC tensor (and those related by $C_3$), $\sigma_{xxz}^S$ transformed in the same way as $\sigma_{xy}^A$, whereas those with all fields in plane: $\sigma_{xxx}^S$ and $\sigma_{yyy}^S$, transform differently and thus provide access to different symmetry information. These planar components are the main focus of this work.

Taking an in-plane magnetic field along $x$ direction the general expression in Eq. \eqref{semiclassicalmain} can be casted into
\begin{align}
\sigma^{S}_{xxx} =\frac{\tau e^2}{\hbar}  \int_{n\bs k} & \Big[  v^x_n \left(  \partial_x  m^x_n- e v^x_n\Omega^x_n -ev^y_n \Omega^y_n \right) \nonumber \\ &- v^{xx}_n m^x_n  \Big] f'_n, \label{semiclassicalxxx}
\end{align}
which explicitly depends on in-plane quantities, namely the Berry curvature and the OMM. 

However, a direct evaluation of Eq. \eqref{semiclassicalxxx} is hindered by the quasi-two-dimensional nature of the layered materials studied in this work, which demands a careful treatment of the bounded $z$-direction. This can be achieved by taking the molecular limit of the bulk expressions for each of its terms. While there are already known expressions for the in-plane Berry curvature and the OMM~\cite{Wang20Optical,Antebi22,Stauber18,Zheng24,Ghorai25}, namely, 
\begin{align}
    \Omega^i_{n} = 2\epsilon^{i j}\text{Im}\bigg[\sum_{q\neq n}r^j_{nq}r^z_{qn}\bigg],
\end{align}
and
\begin{align}
m^i_{n}  = e\epsilon^{ij}\text{Re} \Big[ \sum_{q\neq n} v^j_{nq} r_{qn}^z \Big], \label{OMMlgmain}
\end{align}
respectively, with $i,j \in\{x,y\}$, the in-plane $k$-derivative of the OMM $\partial_im_{j,n}$ is missing in literature. In this light, by using an origin-independent picture\cite{Pozo23} within the length gauge formalism\cite{Aversa95}, we derive an additional sum rule for this quantity which reads
\begin{align}
    \partial_im_{j,n}&= -\frac{e}{\hbar}\epsilon^{jk}\text{Im} \Big[ \sum_{q\neq n} \Delta^j_{nq}r_{nq}^k r^z_{qn} \nonumber \\
    &+ \epsilon_{nq}[r^k_{nq} r^z_{qn;j}+r^k_{nq;j  }r^z_{qn}] \Big].  
\end{align}
This expression, needed to apply the semiclassical formalism for planar responses in quasi-two dimensions, constitutes a key result of this work. There $\epsilon_{nm} = \epsilon_n - \epsilon_m$, $\Delta_{nq}^i = \hbar (v_n^i-v_q^i)$ and $r^i_{nm;j}$ denotes the covariant derivative for the $i$th position operator component along the $j$th direction. Note that the covariant derivatives $r^i_{nm;j}$ with $i \in \{x, y\}$ and $r^z_{nm;j}$ have therefore different expressions. A complete derivation is presented in Appendix~\ref{lengthgaugeexpressions}.

\section{In-plane LMC in Twisted Bilayer Graphene \label{OMRTBG}}
As the most representative example of a class II layered material, i.e., in which VP can be probed via LMC but not through the AHE (see Table~\ref{tab:1}), we now discuss the LMC response in TBG \cite{Bistritzer:PNAS11,dosSantos:PRB12}. This system was the first one where moire-induced topological flat bands were found to give rise to a wide variety of correlated phenomena as a function of filling $\nu$\cite{Cao2018Insulators,Lu19,Kerelsky19,Xie19,Stepanov21,Sharpe19,Serlin20,Cao21Nematicity,Tseng22}, including nematic\cite{Xie:PRB21,Liu:PRR21} and stripe phases\cite{Xie:PRB21,Xie:PRB23}, symmetry-preserving Mott or Kondo\cite{Chou:PRL23,Hu:PRL23,Zhou:PRB24,Rai:PRX24} states, and unconventional superconductivity~\cite{Cao18SC,Yankowitz19,Oh21}. At low temperatures, the correlated phase diagram is governed by two classes of spontaneous symmetry broken (SSB) phases: valley-coherent and (valley-spin) isospin polarized states, whose competition is strongly influenced by strain and substrate effects\cite{Bultinck20, Kwan21,Zhang20BN, Xie20,Bultinck:PRL20,Liu21}. In fact, whereas valley coherence is promoted in pristine samples\cite{Bultinck20, Kwan21}, isospin polarization that yields orbital magnetism is favored at low-strain and high-coupling to the substrate\cite{Kwan21, Zhang20BN, Liu21}. Probing VP states in this intricate phase diagram is experimentally challenging \cite{Penaranda:PRB26,Auerbach:NP25}, making LMC a particularly appealing probe of valley polarization.

Importantly, since valley polarization is only stabilized in the presence of a substrate which breaks $C_{2z}$ symmetry, this would mean VP phases only occur in practice in a class III scenario, where AHE is also allowed. Nevertheless, it is still useful to consider TBG as a class II system because LMC and AHE will have different dependence on the substrate coupling, making the two probes of VP complementary rather than equivalent. For instance, if the substrate coupling is spatially non-homogeneous, as discussed in Sec.~\ref{Mosaics}, the AHE can be reduced and even average to zero, while the LMC is unaffected by this. 

This section is structured as follows: In Sec.~\ref{THFM} we present the self-consistent low-energy model for TBG employed for the LMC calculations. A discussion on the spontaneous-symmetry-broken (SSB) states, a symmetry analysis of the allowed transport responses follows in Sec.~\ref{CorrelatedTBG} together with the LMC calculations. Finally, in Sec.~\ref{Mosaics} we explore the influence of a substrate in the LMC and AHE responses taking into account experimentally relevant scenarios.
\begin{figure}[t]
    \centering
    \includegraphics[width=1\linewidth]{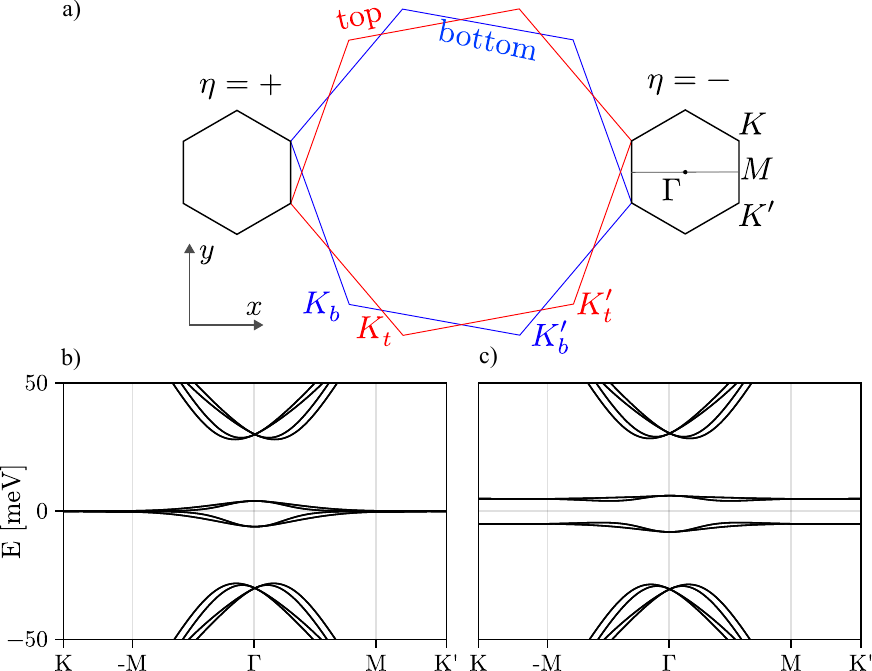}
    \caption{(a) Sketch of the moiré Brillouin zone  of TBG. (b,c) Bandstructures of the non-interacting THFM model of TBG with $\Delta = 0$ (c) and $\Delta \neq 5$ meV (c). Parameters: $M = 5$ meV, $v_\star= -7$ eV\AA , $v'_\star = 2$ eV\AA ,$\gamma = -30$ meV.}
    \label{fig:tbgbands}
\end{figure}
\subsection{The Topological Heavy Fermion Model of TBG}\label{THFM}
The low energy bands of small-angle TBG can be described with the topological heavy fermion model (THFM)\cite{Song:PRL22}. This model preserves the full symmetry of the non-interacting continuum model\cite{Bistritzer:PNAS11} while addressing the mixed localized and itinerant character of charge carriers by considering two types of fermionic operators: the itinerant $c$ and localized $f$ electrons. In this basis, the second-quantized single-particle Hamiltonian reads,
\begin{align}
    \hat H_0 & = \sum_{|\boldsymbol k|<\Lambda_c} \sum_{a, a', \eta, s} \left( H_{aa'}^{(cc,\eta)} (\boldsymbol k) - \mu \delta_{aa'}  \right) \hat c^\dagger_{\boldsymbol k a \eta s} \hat c_{\boldsymbol k a' \eta s} \nonumber \\ &
    - \mu \sum_{\alpha\eta s} \sum_{\bs R} \hat f^\dagger_{\bs R \alpha \eta s} \hat f_{\boldsymbol R \alpha \eta s}  + \frac{1}{\sqrt{N}} \times \\
    & %\times \\&\times
    \sum_{\substack{|\boldsymbol k|<\Lambda_C, \bs R \\\alpha a \eta s}} \left[ e^{i\boldsymbol k \cdot \boldsymbol R - \frac{|\bs{k}|^2 \lambda^2}{2}} H_{\alpha a}^{(fc, \eta)}(\bs k) \hat f^\dagger_{\bs R \alpha \eta s} \hat c_{\boldsymbol{k} a \eta s} + H.c.\right],  \label{H0}\nonumber%
\end{align}
% \interfootnotelinepenalty=10000
where  $\Lambda_c$ is a momentum cutoff, $\hat c^\dagger_{\bs k a \eta s}$ creates a $c$-electron with momentum $\bs k$ in band $a \in \{ 1,2,3,4 \}$, valley $\eta$, and spin $s$, while $\hat f^\dagger_{\bs R \alpha \eta s}$ creates an $f$-electron at position $\bs R$ at orbital $\alpha \in \{1,2\}$ at valley $\eta$ and spin $s$. $\mu$ denotes the chemical potential, $N$ the number of unit cells, and $\lambda = 0.3375 a_M$ a dampening factor accounting for the spread of the $f$ orbitals, and proportional to the moiré lattice constant $a_M$.
\begin{align}
    H^{(cc, \eta)} = \begin{pmatrix}
        0_{2\times2} & v_\star (\eta k_x \sigma_0 + ik_y\sigma_z)\\
        v_\star (\eta k_x \sigma_0 - ik_y\sigma_z) & M \sigma_x
    \end{pmatrix},
\end{align}
denotes the coupling between the $c$ degrees of freedom with $\sigma_{i}$ with $i\in \{0, x,y,z\}$ the Pauli matrices acting on orbital space.
\begin{align}
    H^{(fc, \eta)}(\bs k) = [\gamma \sigma_0 + v_\star' (\eta k_x \sigma_x + k_y \sigma_y), 0_{2\times 2}],
\end{align}
captures the hybridization between $f$ and $c$ degrees of freedom. $M$, $v_\star$, $v_\star'$, and $\gamma$ are determined from the continuum model parameters\cite{Bistritzer:PNAS11,dosSantos:PRB12}. A sketch of the reduced Brillouin zone as well as the single-particle bandstructure at a twist angle $\theta = 1.05^\circ$ are shown in Fig. \ref{fig:tbgbands}a and \ref{fig:tbgbands}b, respectively. 

Accounting for the electron-electron interaction and substrate effects the total Hamiltonian can be written as:
\begin{align}
    \hat H = \hat H_0+\hat H_S+\hat H_I, \label{HITBG}
\end{align}
where $\hat H_S$ and $\hat H_I$ are the corresponding terms, respectively, projected into the $f-c$ basis. In our calculations, we take
\begin{align}
    \hat H_{\text{S}} &= \sum_{\substack{\bs R\\ \alpha \alpha' s \eta}}  \Delta  \hat f^\dagger_{\bs R \alpha s \eta} \sigma^z_{\alpha \alpha'}\hat f_{\bs R \alpha' s \eta}, 
    \label{subs}
\end{align}
corresponding to a substrate-induced sublattice potential with amplitude $\Delta$ that transforms as a $B_2$ irreducible representation of TBG point group. 
This potential reflects the alignment with an hBN substrate, which breaks $C_{2z}T$ while preserving the  $SU(2)_K\times SU(2)_{K'}$ symmetry of independent spin rotations at each valley in Eq.~\eqref{H0}. The corresponding bandstructure features a gap opening at the $K$ and $K'$ points as shown in Fig. \ref{fig:tbgbands}b and c, with $\Delta = 0$ and $\Delta \neq 0$, respectively. As a result, a Chern number given by the product of the sublattice and valley eigenvalue is endowed to each of the four energy-degenerate flat-bands.

The electron-electron interaction is treated under a Hartree-Fock decoupling of the generalized Anderson model that emerges from the projection procedure into the THFM degrees of freedom\cite{Song:PRL22, Penaranda:PRB26}, it reads \begin{align}
H_I \approx \hat H_U + \hat H_J, \label{HITBG}
\end{align}
where
\begin{align}
    \hat H_U & = - \frac{N U}{2} \left( \nu_f^2 + 8\nu_f - \text{Tr}\left[ O^f O^f\right]\right) +U\sum_{\bs R} \sum_{\substack{\alpha s \eta \\ 
    \alpha' s' \eta'}} \nonumber\\
    &\left[ (\nu_f + \frac{1}{2}) \delta_{\alpha \alpha'}\delta_{s s'} \delta_{\eta \eta'} -  O^f_{\alpha s \eta, \alpha' s'\eta'} \right] \hat f^\dagger_{\bs R \alpha' s' \eta' }  \hat f_{\bs R \alpha s \eta}
    \label{HU}
\end{align}
is a local density-density interaction between the $f$ electrons parametrized by $U$ and
\begin{align}
    \hat H_J &= -\frac{J}{2} \sum_{|\bs k|<\Lambda_C} \sum_{\substack{\alpha s \eta \\ \alpha' s' \eta'}} \Bigg[(\eta \eta' + (-1)^{\alpha+\alpha'}) \times \nonumber \\
    & \times \left(O^f_{\alpha \eta s, \alpha' \eta' s'} -\frac{1}{2} \delta_{\alpha \alpha'} \delta_{s s'} \delta_{\eta \eta'}\right)\Bigg] \hat c_{\bs k, \alpha'+2, \eta' s'}^\dagger \hat c_{\bs k, \alpha+2, \eta s},
    \label{HJ}
\end{align}
the exchange Hamiltonian between the $f$ and the $c$ electrons in orbitals $a = \{3,4\}$  with exchange constant $J$. $O^f_{\alpha s \eta, \alpha' s' \eta'} = \langle \psi | \hat f^\dagger_{\bs R \alpha \eta s} \hat f_{\bs R \alpha' \eta' s'}|\psi\rangle$ denotes the density matrix of the $f$ electrons in terms of the interacting mean-field ground state $\psi$, and $\nu_f = \text{Tr}[O^f] - 4$.

\begin{figure}[t]
    \centering
    \includegraphics[width=1\linewidth]{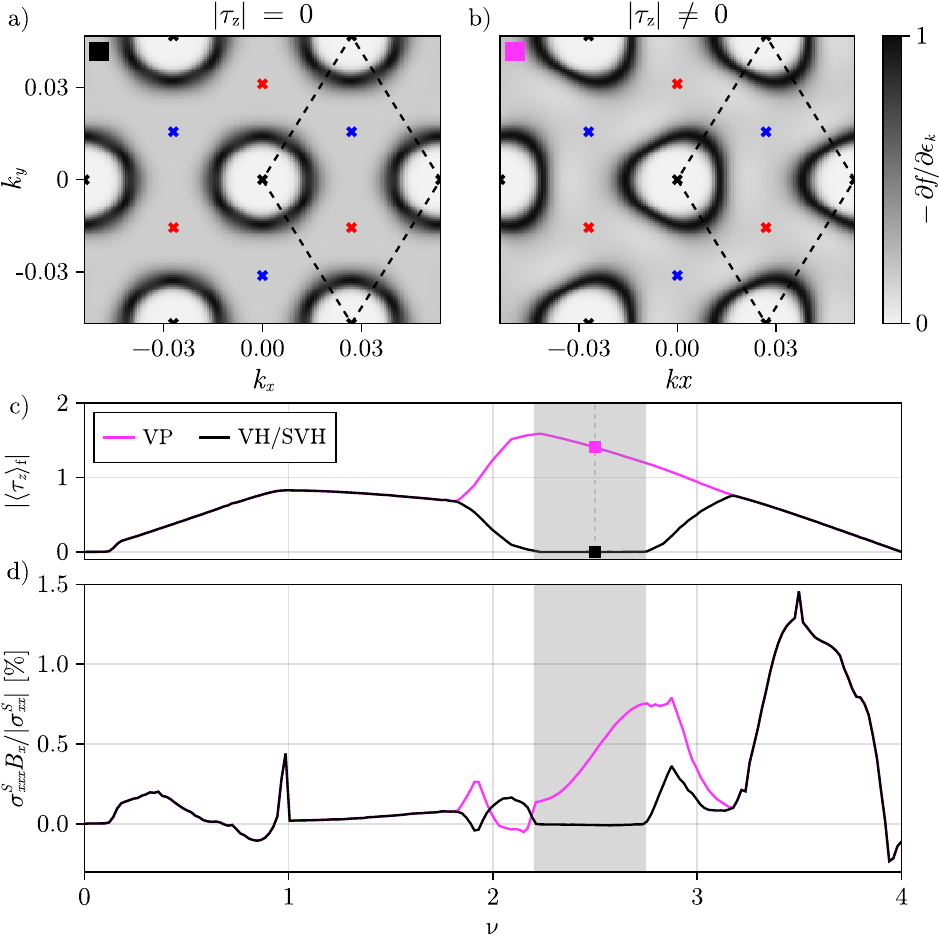}
    \caption{LMC in the interacting SSB states of TBG on a substrate ($\Delta \neq 0$). (a,b) $k$-resolved Fermi surfaces at $\nu = 2.5$ showing two energetically-degenerate ground states of the Hartre-Fock THFM featuring preserved (a) and spontaneously-broken $C_{2y}$ owing to valley polarization (b). (c) Evolution of valley polarization as a function of filling for the two degenerate solutions. Black and pink denote the absence or presence of valley polarization, and the squares indicate $\nu$ values at which the Fermi surfaces in (a,b) were computed. (d) Ratio in percentage of $\sigma^{S}_{xxx} B_x /\sigma^S_{xx}$ as a function of filling. The gray shadow in (c,d) delimitates a filling region with exact valley symmetry for one of the two degenerate solutions (in black) associated with a  symmetry-forbidden LMC response (d). Parameters: $U = 5$ meV, $J=1$ meV, $\Delta = 2.5$ meV, $B_x = 10$ T, $T = 1$ K in (a,b).}
    \label{fig:tbginteracting}
\end{figure}

\subsection{LMC in self-consistent isospin polarized states}\label{CorrelatedTBG}

The $SU(2)_K\times SU(2)_{K'}$ symmetry of the mean field decoupling in Eq.~\eqref{HITBG} imposes a degeneracy among the different configurations of isospin polarization available at a given filling. Whereas odd-integer fillings yield spin and valley polarized ground-states with $|C| = 1$, at $\nu = 2$ several ground-states are energetically degenerate: TRS broken VP state ($|C| = 2$) featuring AHE, and two TRS preserving states:
a ferromagnetic valley Hall (VH) or a spin-valley Hall (SVH) state, corresponding to the equal population of the two valleys and opposite (VH) or same (SVH) spins, both with $|C|=0$\cite{Liu21, Kang19,Xie21}. Supporting evidence for these theoretical predictions was found in transport experiments systematically observing AHE at $\nu = 1$\cite{Stepanov21,Sharpe19,Serlin20,zhang24HoleSide}, but only under specific circumstances at $\nu = 2$\cite{Tseng22}, showcasing the close competition between VP and VH/VSH phases in this filling window.

The self-consistent solutions of the interacting Hamiltonian in Eq. \eqref{HITBG} are presented in Fig. \ref{fig:tbginteracting}, with panels (a,b) displaying the Fermi surfaces of two degenerate ground states: a SVH/VH state that preserves TRS and a VP state that breaks it at $\nu = 2.5$, in (a) and (b), respectively. The breakdown of TRS in (b) manifests as a Fermi surface with broken $C_{2y}$ symmetry. Furthermore, Fig. \ref{fig:tbginteracting}c shows the expectation value of the VP operator $|\langle  \tau_z\rangle|$ as a function filling for the three distinct filling sequences resulting in VH/VSH (black), and VP (pink) phases. The isospin polarization of the two states in (a) and (b) corresponds to those of the black and pink lines at $\nu = 2.5$, respectively. We see how preserved TRS is only realized roughly in the filling region between $\nu = 2$ and $\nu = 3$ for the VH/SVH phase. The degeneracy between VH/VSH and VP can only be broken by extra interactions like a Hund's coupling, which we do not consider here. Hence we simply report the predictions for the two degenerate choices.

\begin{figure}[t]
    \centering
    \includegraphics[width=1\linewidth]{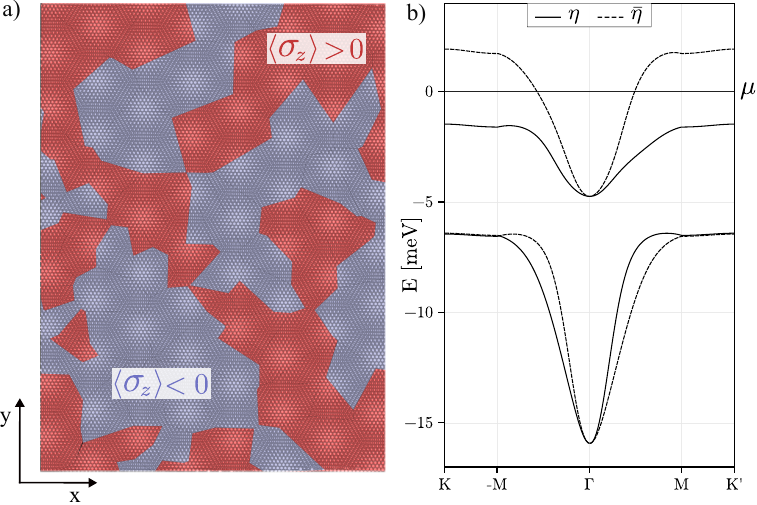}
    \caption{ (a) Illustration of the real space Chern mosaic profile in TBG owing to random sublattice polarization in a ground state with spontaneous valley polarization. $\langle \sigma_z \rangle(x,y)$ is overlaid on top of the moiré lattice with different colors indicating opposite signs. (b) Self-consistent bands of a valley-polarized ground state at $\nu = 2.5$ in the presence of a sublattice potential $\Delta\neq 0$ of arbitrary sign (same parameters as in Fig 3b). Different linestyles indicate different valleys. This regime features a finite LMC response, while the QAHE averages to zero. Same parameters as in Fig.~\ref{fig:tbginteracting}.}
    \label{fig:Mosaic}
\end{figure}
Having obtained the energy-degenerate self-consistent solutions as a function of filling, we now compute $\sigma_{xxx}^{S}$ at $T=1K$, using the quasi-two-dimensional length-gauge formalism derived in Sec.~\ref{LGFormalism} (see Appendix \ref{App:RzinTBG} for details on the implementation). In this section, we neglect the spin contribution to the magnetic moment, which leads to very small corrections to the orbital LMC. This is discussed in more detail in Sec.~\ref{SpinSec}. 

Fig.~\ref{fig:tbginteracting}d  shows the evolution of the ratio in percentage of $\sigma_{xxx}^{S} B_x$ and the longitudinal Drude conductivity $\sigma^S_{xx}$ (see Eq.~\eqref{DRUDE}) for the VH/SVH state in black and the VP state in pink. The gray area delimitates the filling region at which TRS is an exact symmetry in the VH/SVH phase, which therefore leads to a zero LMC response. As we see, the LMC in the VP phase reaches around $ 1\%$ of the total linear response $\sigma^{S}_{xx} + \sigma^{S}_{xxx} B_x$ at $B_x = 10T$. While this value is not large, it is definitely resolvable within experimental capabilities and it is therefore faithful indicator of spontaneous VP in TBG. 

\subsection{Experimental implications for mosaic states\label{Mosaics}}

As explained in the introduction to this section, TBG is a class II system only in the absence of a substrate. Once the substrate-induced potential $\Delta \neq 0$ is included, which breaks $C_{2z}$ and $C_{2x}$, the AHE becomes allowed and TBG becomes class III. Nevertheless, LMC can only depend quadratically on $\Delta$, while the AHE is proportional to $\Delta$ and hence switches sign when $\Delta$ does.

This distinction is important because in experiments the substrate coupling has a complex spatial profile, simulated with constant $\Delta$ only as a first approximation. Local sensing measurements in TBG\cite{Grover22} have in fact found a non-uniform magnetization profile across the sample, resulting in a so-called Chern mosaic\cite{BhattacharjeeA26} with domains carrying opposite Chern numbers. This could be the result of either a constant substrate coupling with domains with opposite valley polarization, or as argued in Ref. \cite{Grover22}, more likely a result of constant valley polarization and spatially varying substrate coupling $\Delta(x,y)$ that randomly changes sign. This second interpretation is more natural because the sign of the substrate coupling depends on the local alignment with hBN and moiré reconstruction.

This situation is schematically illustrated in Fig. \ref{fig:Mosaic}a where a random sublattice potential is overlaid on top of the TBG lattice, and different colors encode different signs of the sublattice polarization $\langle\sigma_z\rangle(x,y)$, while a homogeneous valley polarization $\langle \tau_z\rangle(x,y) = \langle \tau_z \rangle$ is assumed across the sample. Rather than the insulating case, for this example we consider a metallic case with non-quantized $\sigma_{xy}^{A}$ and a Fermi surface, as shown in \ref{fig:Mosaic}b. In such a situation, although $\sigma_{xy}^{A}$ of each individual magnetization domain is symmetry allowed due to the substrate-induced symmetry breaking, the total contribution $\sigma_{xy}^{A}\propto\langle\tau_z\rangle \langle\sigma_z\rangle$ averages out to zero. Conversely, the different symmetry of $\sigma_{xxx}^{S} \propto \langle \tau_z \rangle$ ensures it is finite even in this scenario. Hence, we argue $\sigma^{S}_{xxx}$ is not only finite and measurable in valley polarized TBG, but it is also effectively the unique faithful probe of valley polarization in this type of experiment.

\begin{figure}[t]
    \includegraphics[width=1\linewidth]{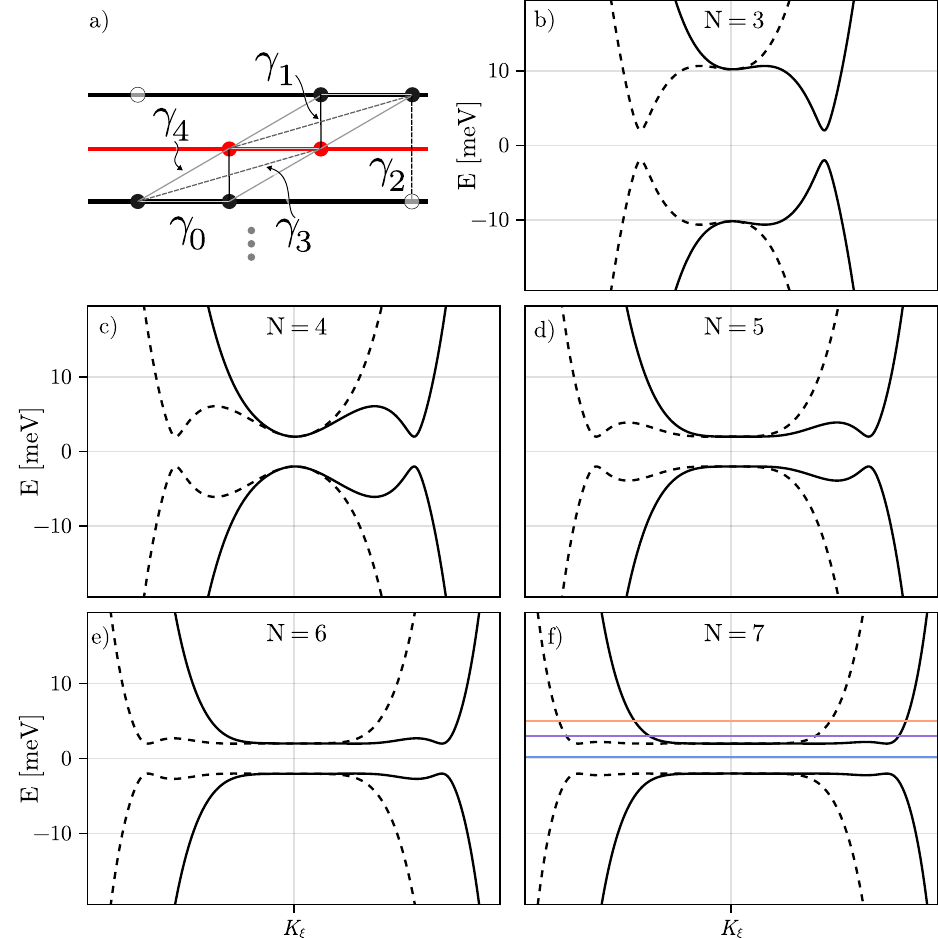}
    \caption{Non-interacting Koshino-McCann model. (a) Sketch of the device consisting of N-layers ABC stacked with intra and interlayer hopping parameters $\gamma_{i}$ with $i\in
    \{1,2,3,4\}$. (b-f) Bandstructures for $N=\{3,4,5,6,7\}$ layers with: $\gamma_0 = 3.16$ eV, $\gamma_1 = 0.39$ eV, $\gamma_2=-0.02 eV$, $\gamma_3=0.315$ eV, and $\gamma_4 = 0$. Solid and dashed lines correspond to the two different valleys at $K_{\pm}$. The $E_z$-induced potential $\Delta = 2$ meV gaps the spectrum.}
    \label{bandsRHG}
\end{figure}
\section{In-plane LMC in Rhombohedral N-layer Graphene \label{OMRRNLG}}

In this section, we consider N-layer RG as a representative system from the class III in Table \ref{tab:1}, where both LMC and AHE are allowed.  %to find that LMC offers a general protocol to extract valuable information regarding the valley polarization of its ground states.
TBG and N-layer RG share the key features of non-trivial band topology and a high density of states in narrow low-energy bands, which also gives rise in RG to a rich phase diagram characterized by multiple interaction-driven symmetry broken phases\cite{Zhou:21N, Zhou:21N2,Liu:NN24, Sha:S24, Choi:N25,Han:N23, Han:NN24,Qin:A26, Deng:A25, Zheng:PRL25,Zhou:NC24,Han:2025, Qin:A26}, including as well, isospin polarized phases and unconventional superconductors. The layer number $N$ in N-layer RG provides an additional knob that controls the bandwidth and quantum geometry of the low-energy bands, allowing to tune interaction effects across multilayer thicknesses and to explore different ways of breaking the valley-spin isospin degeneracy.

This section is organized as follows: Secs.~\ref{KMmodelSec} and \ref{IntRG} contains the non-interacting and interacting models, respectively, used in our calculations with Sec.~\ref{CorrRG} displaying a classification of its correlated ground states. Our calculations within the semiclassical Eq.~\eqref{semiclassicalmain} are discussed in Secs.~\ref{LMCnonintRG} and ~\ref{LMCintRG}, in the absence and presence of Hartree-Fock interactions, respectively. Finally, a study of the spin contribution to $\sigma_{xxx}^S$ is presented in Sec.~\ref{SpinSec}.

\subsection{The Koshino-McCann model}\label{KMmodelSec}
LMC in N-layer RG is computed on the basis of the Koshino-McCann model\cite{Koshino:PRB09}. This continuum model around $K$ and $K'$ valleys describes the two lowest-energy bands which are generated by the non-dimer sites in a graphene multilayer ABC-stack with arbitrary $N\geq 3$, i.e., those generated by the $A_1$ and $B_N$ sites that lie on the top and bottom layers, respectively. The single-particle Hamiltonian for valley $\xi$ reads
\begin{align}
    H(p) = \begin{pmatrix} \Delta (E_z) & X(p)\\
    X^\dagger(p) & -\Delta(E_z)
    \end{pmatrix} + \frac{2v_0v_4p^2}{\gamma_1} \begin{pmatrix}
        1 & 0 \\ 0 & 1
    \end{pmatrix}, \label{KMmodel}
\end{align}
where $p = \hbar|\bs k|$ is the momentum measured with respect to $K (K')$, \begin{align}
    X(p)= \sum_{\{n_1,n_2,n_3\}}  &\frac{(n_1+n_2+n_3)!}{n_1!n_2!n_3!} \frac{1}{(-\gamma_1)^{n_1+n_2+n_3-1}} \times \nonumber \\
    & (v_0p e^{i\xi \phi})^{n_1} (v_3p e^{-i\xi \phi})^{n_2} \left(\frac{\gamma_2 }{2}\right)^{n_3}, \label{Xp}
\end{align} $v_i = (\sqrt{3}/2) a \gamma_i/\hbar$, $a$ is the graphene lattice constant, $n_i$ includes all positive integers satisfying $n_1+n_2+n_3 = N$. $\gamma_i$ denotes the Slonczewski-Weiss-McClure parameterization of the tight-binding couplings in bulk graphite: $\gamma_0 = 3.16\ \text{eV}$ is the intralayer coupling, $\gamma_1= 0.39\ \text{eV}$ the nearest-layer coupling between sites ($B1$-$A2$ and $B2$-$A3$),  $\gamma_2=-0.020\ \text{eV}$ the next-nearest-layer coupling between sites $A1$ and $B3$, and $\gamma_3=0.315\ \text{eV}$ and $\gamma_4=0.044\ \text{eV}$ as the nearest-layer couplings between different sublattices, and the same sublattice, respectively. These are schematically shown in Fig. \ref{bandsRHG}a.

Furthermore, the phase diagram of N-layer RG is strongly affected by an externally-applied out-of-plane electric field $E_z$ that opens a gap in the non-interacting spectrum at charge neutrality. Its effect is parametrized in Eq. \eqref{KMmodel} by $\Delta(E_z) =eE_z d/2$, where $d$ is the separation between the first and outermost layer. A key property of relevance for the LMC response is the flattening of the electron and hole bands with increasing $N$. These spectral features are shown in Fig. \ref{bandsRHG}b-f for increasing layer index $N$ with $\Delta(E_z) \neq 0$. 

\subsection{Interactions - The SU(2) rigid band model}\label{IntRG}
Electronic correlations are introduced in the McCann-Koshino model by means of a local Hartree interaction\cite{Zhou:21N} given by
\begin{align}
    V_{int} = \frac{UA}{2} \sum_{\alpha \neq \beta} n_\alpha n_\beta + J_H A (n_1 - n_3)(n_2 - n_4), \label{SU2int}
\end{align}
where $A$ is the unit-cell area, $n_\alpha$ with $\alpha \in \{1,2,3,4\}$, are the expectation values of the electron densities of the four isospin flavors: $1 = \{K, \uparrow\}$, $2 = \{K', \uparrow\}$, $3 = \{K, \downarrow\}$, and $4 = \{K', \downarrow\}$. The first term in Eq. \eqref{SU2int} accounts for a $SU4$-preserving screened Coulomb interaction parametrized by $U$ that favors quarter-metallicity. Whereas, the second term explicitly breaks the intrinsic isospin symmetry of the Koshino-McCann model and makes the Hamiltonian $SU(2)$ symmetric. This term corresponds to a ferromagnetic Hund's interaction with negative coupling constant $J_H$.\\

The local character of the interaction turns the problem into a rigid-band model where the effect of the interaction merely shifts in energy each spin and valley flavor separately without altering their dispersions. The ground states can be obtained by minimizing the grand energy potential 
\begin{align}
    \frac{\Phi}{A} = \sum_{\alpha} E(\mu_\alpha) + V_{\text{int}} - \mu \sum_{\alpha} n(\mu_\alpha), \label{SCintRG}
\end{align}
written in terms of $\mu_\alpha$, the chemical potentials for each isospin species. Note that by construction, this model neglects intervalley coherent states.

\subsection{Correlated states}\label{CorrRG}

We restrict our analysis to the intermediate interaction strengths, where the rigid-band model remains a valid approximation. This regime characterized by isospin polarized SSB phases is in agreement with experimental observations that have found orbital magnetism as an ubiquitous feature in RG multilayers regardless of $N$. Indeed, it has been reported in trilayers\cite{Zhou:21N, Zhou:21N2}, tetralayers\cite{Liu:NN24, Sha:S24, Choi:N25}, pentalayers\cite{Han:N23, Han:NN24}, hexalayers\cite{Qin:A26, Deng:A25, Zheng:PRL25}, or heptalayers\cite{Zhou:NC24}. 
\begin{figure}[t]
    \includegraphics[width=1\linewidth]{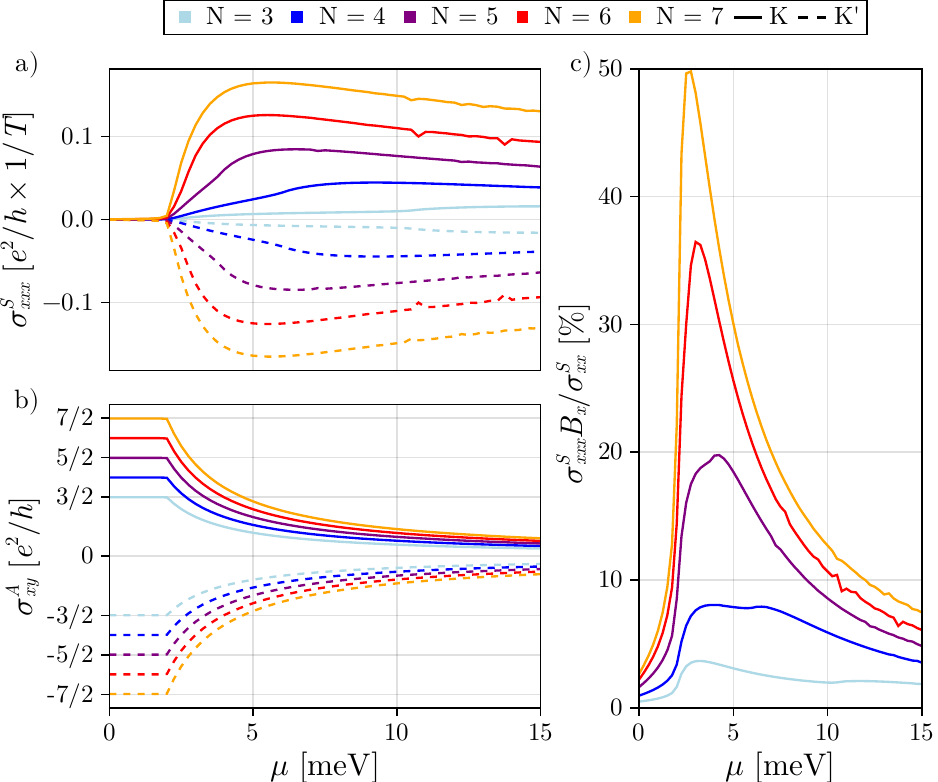}
    \caption{LMC for the non-interacting bands of N-ABC rhombohedral graphene. (a) $\sigma^{S}_{xxx}$ and (b) $\sigma_{xy}$ versus $\mu$ for the non-interacting bands of a given spin, where colors indicate the number of layers of the N-ABC RHG stack, and solid and dashed lines correspond to different valleys $K$ and $K'$. (c) $\sigma^{S}_{xxx} B/\sigma^S_{xx}$ in percentage as a function of $\mu$.}
    \label{fig:noninteractingrnlgomr}
\end{figure}

In the absence of Hund's interaction ($J_H = 0$), SSB phases in RG similar to TBG can be classified by its particular spin-valley isospin polarization within the energetically-degenerate $SU(4)$ manifold. Therefore, by doping the system and at some finite $E_z$, the following  ground states may alternate: symmetric $SU(4)$-preserving phases, quarter metals (QM) around odd-fillings corresponding to the occupation of an odd number of bands, necessarily resulting in orbital magnetism, and half-metal (HM) phases around $\nu = \pm 2$, that may or may not exhibit orbital magnetism depending on whether the two spins within a single valley (VPHM), or the two valleys of a given spin (SPHM), are equally populated, respectively.

The intervalley Hund's interaction in Eq.~\eqref{SU2int}, which lowers the Koshino-McCann $SU(4)$ symmetry to $SU(2)$,  is responsible for establishing a hierarchy among these otherwise-degenerate HM states. The ferromagnetic coupling constant ($J_H<0$) considered in this work, and in line with experimental estimates in Ref.~\cite{Auerbach:NP25}, favor SPHM over VPHM.

\subsection{Non-interacting in-plane LMC calculations}\label{LMCnonintRG}

As a proof of concept, we begin the analysis of the in-plane LMC response in N-layer RG by analyzing $\sigma_{xxx}^{S}$ for a single degenerate band in the non-interacting Hamiltonian of Eq.~\eqref{KMmodel}. This minimal model describes a correlation-induced QM regime in which only a single decoupled isospin flavor contributes to the LMC. We employ once more the semiclassical formalism in Eq. \eqref{fullsemiclassical} and the sum rules derived in Sec.~\ref{LGFormalism} to analyze its response.

The results are presented in Fig. \ref{fig:noninteractingrnlgomr}. The dependency of $\sigma_{xxx}^{S}$ and $\sigma_{xy}^{A}$ (see Eq~\eqref{HALL}) on $\mu$ is shown in panels (a) and (b), respectively, for each of the bandstructures in Fig. \ref{bandsRHG}b-f, with different colors denoting different number of layers. Since the LMC is a Fermi surface property, $\sigma_{xxx}^{S}$ is zero inside the $E_z$-induced gap, it increases rapidly at the gap edge, and exhibits a slow decay with doping as we depart from it. This behavior contrasts with the anomalous Hall response in panel (b) computed using Eq.~\eqref{j0}. When summed over the two valleys it yields quantization inside the gap, owing to an integrated Berry phase equal to $N\xi \pi$,  but rapidly decreases once the electron-band starts being populated. In Fig.~\ref{fig:noninteractingrnlgomr}c a  comparison between the ratio expressed as a percentage between the LMC at $B=10T$ and the Drude conductivity in the $x$-direction $\sigma_{xx}^{S}$ is presented. Remarkably, we find that the LMC constitutes a large percentage of the total response  to order $B^2$ in extended chemical potential regions, displaying increasing ratios with increasing number of layers, reaching the $50\%$ for $N = 7$ ($\sim 1.5 e^2/h$ in absolute value) at $B=10T$.

\begin{figure}[t]
    \centering
    \includegraphics[width=1\linewidth]{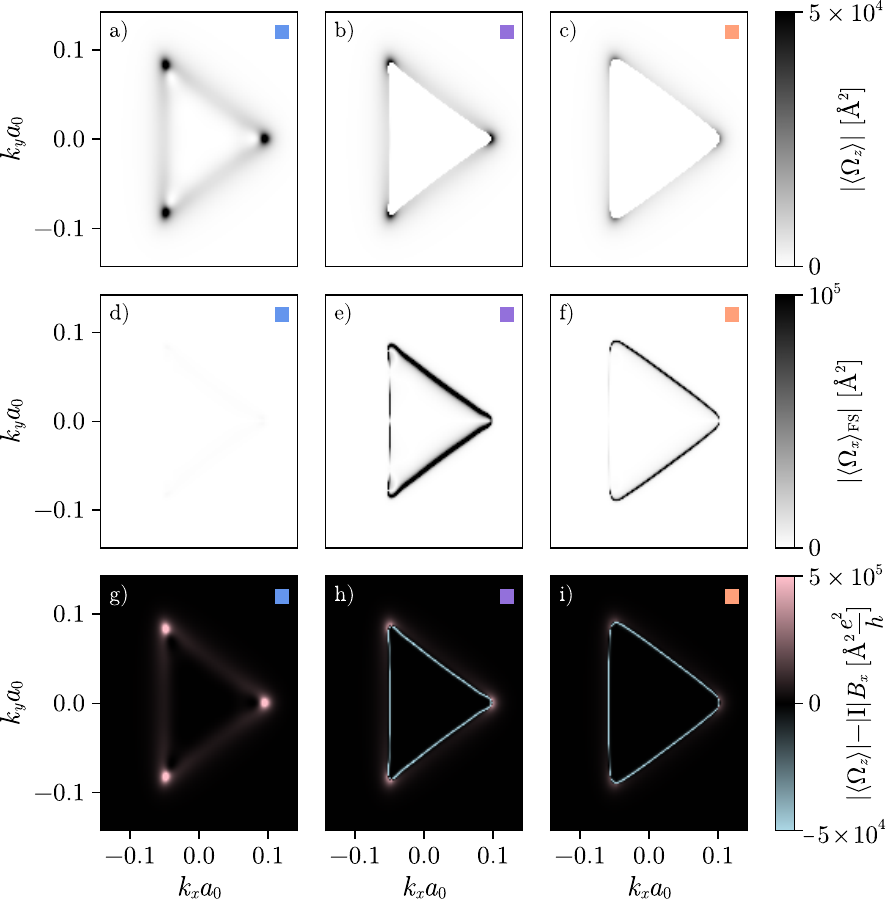}
    \caption{Momentum-resolved quantities involved in $\sigma^{S}_{xxx}$ as a function of filling for heptalayer RHG. (a-c) Expected value of the out-of-plane Berry curvature and (d-f) in-plane Berry curvature at the Fermi surface at the three different $\mu$ values in Fig. \ref{bandsRHG}(f). (g-i) Difference between $\sigma^A_{xy}$ and $\sigma_{xxx}^{S}$ integrands  at $B = 10$ T, where $I$ denotes the integrand of Eq.~\eqref{fullsemiclassical}. A faster decay of $|\langle\Omega_{z}\rangle|$ with filling compared to $|\langle \Omega_{x}\rangle_{FS}|$ explains the competition between $\sigma_{xxx}^{S}$ and $\sigma_{xy}$ as the leading TRS-breaking probe away from the van Hove. Same parameters as in Fig.~\ref{fig:noninteractingrnlgomr}.}
    \label{fig:kresolved}
\end{figure}

The different functional dependencies of the two responses together with the large values found for the LMC, suggests that under specific circumstances the largest transport probe of valley polarization is indeed $\sigma_{xxx}^{S}$. 
To clarify the mechanism we now provide a momentum-resolved study of the two responses. Fig.~\ref{fig:kresolved} compares the integrand of the semiclassical expressions for the LMC and the AHE resolved in momentum. Where each column is associated with different $\mu$ values marked in Fig.~\ref{fig:kresolved} by colored horizontal lines, (a-c) show the expected value of the out-of-plane Berry curvature, (d-f) the in-plane Berry curvature at the Fermi surface, and (g-i) the difference between $\sigma_{xy}^{A}$ and $\sigma_{xxx}^{S}$ integrands, namely $|\Omega_z|$ and $|I B_x|$ where $I$ is sum of the integrands in Eq.~\eqref{semiclassical} and Eq.~\eqref{correction}. 
The integrand of $\sigma_{xy}^{A}$ (in Eq.~\eqref{HALL}) is the $z$-component of the Berry curvature and thus a Fermi sea quantity. Upon doping away from the gapped Dirac cones where the Berry curvature is sharply concentrated (a,g), it decays sharply (panels b,c), leaving only a residual contribution in (h,i). In contrast, the in-plane Berry curvature (d-f), which enters as the integrand of $\sigma_{xxx}^{S}$ (g-i), receives a nearly homogeneous contribution from the entire Fermi surface and tracks the density of states, peaking at the van Hove singularity and decaying slowly as the doping moves away.

\begin{figure*}[t]
    \centering
    \includegraphics[width=1\linewidth]{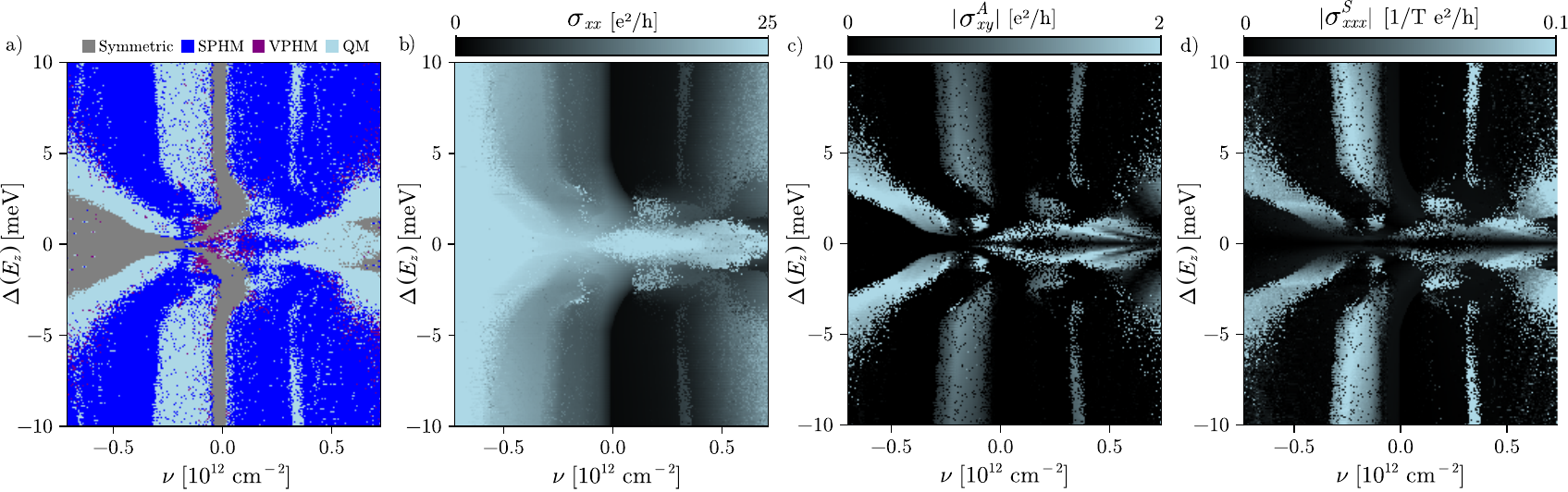}
    \caption{LMC in the interacting Koshino-McCann model of heptalayer graphene. (a) Phase diagram filling vs $E_z$ in the presence of a $SU(2)$ interaction that preserves the valley. Symmetric, spin polarized and valley polarized half metals and (three) quarter phases are depicted in gray, blue, purple, and cyan, respectively. (b) $\sigma^S_{xx}$, (c) $|\sigma^A_{xy}|$, and (d) orbital contribution of $|\sigma_{xxx}^{S}|$ for each point $(\nu, E_z)$ in (a). Parameters: $U = 10$ meV, $J = -4$ meV, the rest of parameters are equal to those of Fig. \ref{bandsRHG}.}
    \label{fig:pdmain}
\end{figure*}

\subsection{Interacting in-plane LMC calculations}\label{LMCintRG}

We now examine whether the intuition developed based on the QM minimal model remains valid when interactions are introduced. This is achieved by computing the in-plane LMC response across the $(n, E_z)$ phase diagram using the interacting self-consistent solutions of Eq.~\ref{SCintRG}. 

We chose in this analysis heptalayer RG, although similar results yield for arbitrary $N$ (see Appendix~\ref{NlayerRGLMC}). 
The corresponding phase diagram is shown in Fig. \ref{fig:pdmain}a. Gray indicates the $SU(4)$ symmetric phase which dominates at charge neutrality and hole dopings at low $E_z$, light blue indicates QM, and dark-blue and purple SPHM and VPHM phases. Hund's interaction results in a predominancy of SPHM over VPHM phases and the particle-hole asymmetry is a direct consequence of $\gamma_4 \neq 0$. The obtained sequence of QM and HM phases is in agreement with theoretical HF studies\cite{Zhou:21N, Parra:A25} using the 2N band model, indicating that the two-band model is adequate to describe the mean-field phenomenology in this intermediate interaction regime. 

The Drude conductivity $\sigma_{xx}^{S}$ is shown in panel (b) acting as an spectral probe that captures the sequential filling of isospin flavors in close resemblance to TBG. The absolute values of $\sigma_{xy}^{A}$ and orbital contribution of $\sigma_{xxx}^{S}$ are shown in panels (c) and (d), respectively. First, we observe that both quantities follow closely the phase diagram, being finite whenever there is VP, i.e., in the QM and in the VPHM phases, and vanishing in the SPHM and symmetric phases. Similar to the non-interacting case in the section above, we find large values of the LMC response of comparable magnitude to the AHE at moderate $B$ fields.

\begin{figure}[t]
    \centering
    \includegraphics[width=1\linewidth]{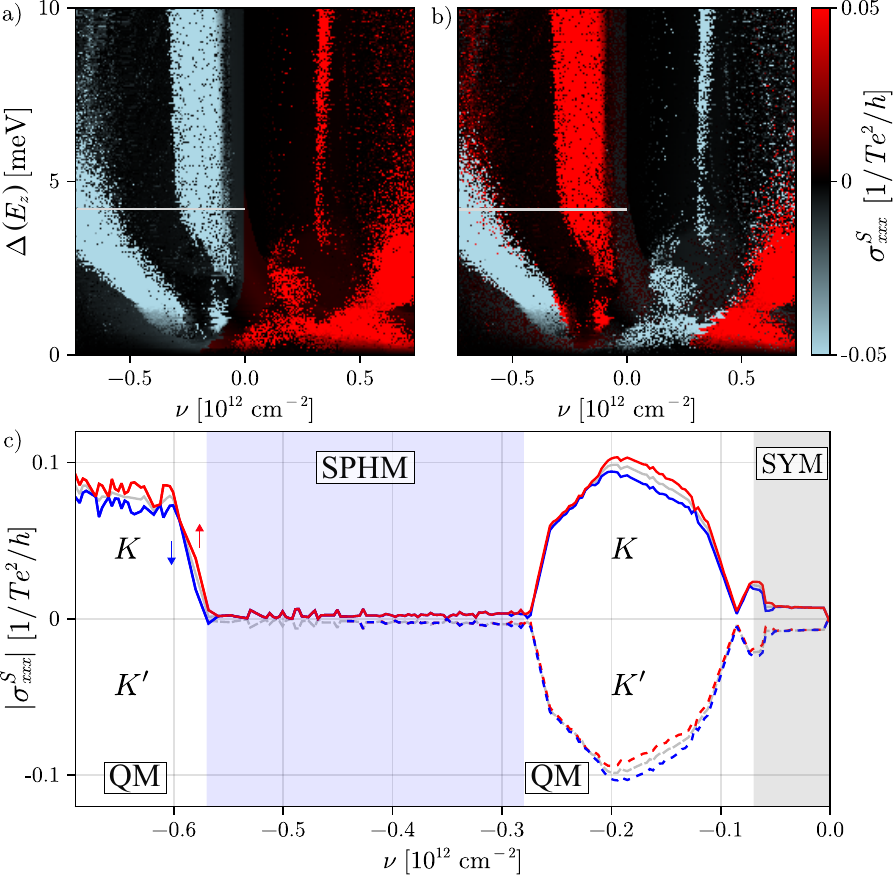}
    \caption{Orbital and spin contributions to $\sigma^{S}_{xxx}$  in heptalayer RHG. Total (orbital+spin) $\sigma_{xxx}^{S}$ for two inequivalent filling sequences in the $SU(2)$-interacting model: (a) $\uparrow K$, $\downarrow K$, $\uparrow K'$, $\downarrow K'$ and (b) $\uparrow K$, $\downarrow K'$, $\uparrow K$, $\downarrow K'$. (c) Total  $\sigma_{xxx}^S$ along the linecuts in panels (a-b). Different colors denote different spins and different linestyles different valley polarizations. Same parameters as in Fig.~  \ref{fig:pdmain}.}
    \label{fig:spinrhg}
\end{figure}

\subsection{Spin contribution to LMC}\label{SpinSec}

Even in the presence of Hund's coupling, there is still an energy degeneracy among isospin polarized ground states. Fig.~\ref{fig:spinrhg} examines the LMC signatures of the two isospin filling sequences that are inequivalent and degenerate under the $SU(2)$ interaction. They differ in the ordering of occupied valley ($K/K'$) and spin $\uparrow/\downarrow$ flavors with doping: $[\uparrow K$, $\downarrow K$, $\uparrow K'$, $\downarrow K']$ and $[\uparrow K$, $\downarrow K'$, $\uparrow K$, $\downarrow K']$. Panels (a) and (b) show the orbital contribution to $\sigma_{xxx}^S$  for each filling sequence, respectively. As already anticipated, the sign of the orbital contribution to $\sigma_{xxx}^S$ at each QM phase is determined by the occupied valley, showing the two filling sequences different sign alternations with doping.

The addition of the Zeeman (spin) contribution to the LMC breaks the spin degeneracy at each valley. To quantify its effect, we plot in panel (c) line-cuts at fixed $\Delta(E_z)$ along filling values corresponding to the horizontal gray lines in (a) and (b). In gray we show the orbital contribution  and in red and blue the total contribution for parallel or antiparallel spins with respect to $B$, where solid and dashed lines refer to the $K$ and $K'$ valleys. While the spin correction to LMC is an order of magnitude smaller than the orbital part, if a precision measurement is feasible as a function of carrier concentration this might be used to unambiguously determine the spin and valley polarization of each QM phase.

\section{Discussion\label{Discussion}} 

In this work, we identified two classes of systems where LMC is a direct probe of spontaneous valley polarization: In class III, this response occurs alongside the AHE, while in class II it represents the only probe of valley polarization because the AHE vanishes by symmetry. So far, the characteristic butterfly resistivity hysteresis (see Fig.~\ref{fig:sketch}b) associated to LMC has been observed mostly in the out of plane geometry alongside the AHE, as for example in TMDs like AB-stacked MoTe2 /WSe2 moiré heterobilayers \cite{Li21} and twisted bilayer MoTe2\cite{Xu23}, as well as in rhombohedral graphene~\cite{Han:S24,Shi:N20,LiA25}. These observations are natural as they all belong to class III, although a word of caution might be appropriate since in some cases such hysteresis has been claimed to originate from thermal effects \cite{Dutta25}. 

A particularly large hysteresis was reported for trilayer RG in Ref. \cite{Lee22} with both in-plane and out of plane fields, an interpreted in terms of a doped layer antiferromagnetic state. However, our analysis and explicit calculations reveal that both AHE and LMC are only consistent with a quarter metal state, which should thus be the one realized in Ref. \cite{Lee22}. Importantly, these experiments were done in suspended samples where only two terminal transport is possible, precluding the observation of the AHE. In this case, LMC remains the only probe of valley polarization despite the fact that this system is nominally in class III. This experiment serves as an example of the power of LMC as a probe of valley polarization. On a more general note, a thorough characterization of isospin polarization in RG multilayers is of great importance as these are the parent states of even more exotic states, for instance, of chiral topological superconductors\cite{Han:2025, Yoon:PRL26,QinA26}, and we believe LMC can become a useful probe in this regard. 

Interestingly, in-plane LMC has also been observed in a class II system, in twisted trilayer graphene at finite displacement field in a metallic region near the superconducting state~\cite{Mukherjee24}, while no clear AHE hysteresis in $B_z$ is observed. In this system (see Table \ref{tab:1}) the AHE is forbidden by $C_{2z}\mathcal{T}$. This observation would thus be consistent with a valley polarized state, as predicted by our theory, and serves as another example where LMC probes such valley polarization even when the AHE vanishes. 

We have also discussed how TBG is an interesting example in class II where formally a valley polarized state leads to LMC without AHE, but in practice the presence of a substrate which breaks $C_{2z}$ is required to stabilize the valley polarized state over competing intervalley coherent ground states. Even if the breaking of $C_{2z}$ would take this system to class III and enable AHE, the symmetry analysis and calculation is still useful to stress the fact that LMC is proportional to valley polarization alone, while AHE is also proportional to the substrate perturbation and hence would average out in a mosaic state where the sign of the substrate perturbation is random. An observation of LMC in TBG in the absence of AHE, which we believe is within experimental reach, would represent a confirmation of our prediction for the mosaic state. 

This type of symmetry analysis is more generally useful when the perturbation enabling AHE can be externally controlled. For example, AB-AB TDBG belongs to class II with the AHE forbidden by an in-plane two-fold axis, which can be broken externally by an applied displacement field. This brings the system to class III, but the AHE is proportional to such field, while the LMC is independent of it. These examples show the usefulness of considering a pristine valley polarized state in class II as a starting point, even if experimentally some symmetries might be broken in practice and bring the system to class III.

In summary, with this work we hope to draw attention to in-plane LMC as a probe of valley polarization in layered heterostructures which is independent of the AHE and establish its relevance for the identification of correlated states in current transport experiments.

\emph{Acknowledgements} - We thank Felix Casanova, Veronika Sunko and Hector Ochoa for insightful discussions. F.~J.~acknowledges funding from Grant PID2021-128760NB0-I00 from the Spanish MCIN/AEI/10.13039/501100011033/FEDER, EU and from a 2024 Leonardo Grant for Scientific Research and Cultural Creation, BBVA Foundation. F.P acknowledges support from a Juan de la Cierva Fellowship (Grant No. JDC2023-051274-I) funded by MICIU/AEI/10.13039/501100011033 and the ESF+.

\appendix 

\section{Semiclassical equations \label{AA}}
In this Appendix we provide a derivation of Eq.~\eqref{semiclassicalmain} employing the semiclassical Boltzmann transport formalism for non-degenerate bands\cite{Sundaram99, Xiao10}.  In the presence of static electric $\bs E$ and  magnetic $\bs B$ fields, the semiclassical expression in $d$ dimensions that describes the current density resulting from the field-induced wave-packet  motion reads
\begin{align}
 \bs j = \sum_n \bs j_n = -e \int_{n\bs k} \frac{\dot{\bs r}_n(\bs k) f_n (\bs k) }{D_n(\bs k)}, \label{Currentdensity}
\end{align}
valid in the $\omega_c \tau \ll 1$ where $\omega_c$ and $\tau$ are the cyclotron frequency and the relaxation time, and where $\int_{n\bs k} \equiv \sum_n \int \frac{d^d k}{(2pi)^d}$, $e>0$, $\dot{\bs r}_n$ is the time derivative of the wave-packet center-of-mass position of the $n$'th band, $D = 1 + (e/\hbar) (\bs B \cdot \bs \Omega_{n\bs k})$, the phase-space density correction due to the static magnetic field\cite{Di:PRL95}, and $f_n$, the distribution function of band $n$ which satisfies the kinetic equation:
\begin{align}
    \frac{\partial f_n (\bs k)}{\partial t} + \dot{\bs k} \cdot \bs \nabla_{\bs k} f_n (\bs k) = - \frac{f_n(\bs k)-f^{(0)}_{n}(\bs k)}{\tau}. \label{Kinetic}
\end{align}
Note that Eq.~\eqref{Kinetic} is written under the field-independent relaxation time ($\tau$) approximation\cite{Xiao10} that neglects the field-induced intra-scattering dynamics\cite{Xiao20} and under the assumption of a spatial homogeneity, i.e., $f_n(\bs r, \bs k) = f_n(\bs k)$, with $f^{(0)}_{n}(\bs k)$ denoting the equilibrium Fermi-Dirac distribution.

An expression for $\dot{\bs r}_n$ can be obtained from the Bloch wave-packet equations of motion. They read
% \begin{align}
%     &\hbar \dot{\bs r}_n = \bs \nabla_{\bs k} \epsilon_{n\bs k} - \hbar \dot{\bs k}_n \times \bs \Omega_{n \bs k}, \nonumber \\
%     &\hbar \dot{\bs k}_n = -e\bs E - e \dot{\bs r}_n \times \bs B, \label{EoM}
% \end{align}
\begin{align}
    &\hbar \dot{\bs r}_n = \frac{1}{\hbar D_{n \bs k}} \left( \bs \nabla_{\bs k} \varepsilon_{n\bs k} + e \bs E \times \bs \Omega_{n \bs k} + \frac{e}{\hbar} \bs B (\bs \Omega_{n\bs k} \cdot \bs \nabla_{\bs k} \varepsilon_{n \bs k}\right), \nonumber \\
    &\hbar \dot{\bs k}_n =  \frac{1}{\hbar D_{n \bs k}}  \left( -e \bs E - \frac{e}{\hbar} \bs \nabla_{\bs k} \varepsilon_{n \bs k} \times \bs B - \frac{e^2}{\hbar} (\bs E \cdot \bs B) \bs \Omega_{n \bs k} \right), \label{EoM}
\end{align}
where $\bs \Omega_{n \bs k } = i \langle \bs \nabla_{\bs k} u_{n \bs k}|\times \bs \nabla_{\bs k} u_{n \bs k } \rangle$ is the Berry curvature with $|u_{n\bs k}\rangle$ denoting the eigenvectors of the crystal Hamiltonian $H|u_{n\bs k}\rangle = \varepsilon^0_{n \bs k} |u_{n \bs k}\rangle$ at $\bs B = \bs 0$, and $\varepsilon_{n \bs k}$ the B-induced bare energy modification: \begin{align}
    \varepsilon_{n \bs k} = \varepsilon^0_{n\bs k} -\bs m_{n\bs k} \cdot \bs B,
\end{align} in terms of the orbital magnetic moment
\begin{equation}
    \bs m_{n\bs k} = -i \frac{e}{2\hbar} \langle \bs \nabla_{\bs k} u_{n \bs k}|\times (H_{\bs k} - \varepsilon^0_{n \bs k}) \bs |\nabla_{\bs k} u_{n \bs k}\rangle. \label{OMM}
\end{equation}
In order to obtain an expression for Eq.~\eqref{Currentdensity} to linear order in the fields, the non-equilibrium distribution of band $n$ can be perturbatively expanded in the weak-$\bs E$-field limit as:
\begin{align}
    f_n = \sum_{i=0}^\infty f^{(i)}_{n}
\end{align}
where $f^{(i)}_{n} \propto E^i$. Plugging Eq.~\eqref{EoM} for $\dot{\bs k}_n$ into Eq.~\eqref{Kinetic}, to first order in $\bs E$ and $\bs B$, such that $f_n \approx f^{(0)}_{n} + f^{(1)}_{n}$, we obtain
% \begin{align}
%     f^{(1)}_{n} = \frac{e\tau}{\hbar D_{\bs k}} &\Bigg[ \left( \bs E + \frac{e}{\hbar} (\bs E \cdot \bs B) \bs \Omega_{n \bs k}  \right) \cdot \nabla_{\bs k} f^{(0)}_{n} \nonumber \\
%     & + \frac{e\tau}{\hbar^2} (\nabla_{\bs k} \varepsilon_{\bs k} \times \bs B ) \cdot \bs \nabla_{\bs k}(\bs E \cdot \bs \nabla_{\bs k} f^{(0)}_{n} ) \Bigg], \label{distribution}
% \end{align}
% Furthermore, keeping Eq. \eqref{distribution} to first order in $\bs B$ yields:
\begin{align}
    f^{(1)}_{n} \approx \frac{e\tau}{\hbar} &\Bigg[ 
    \Big[\hbar \bs E \cdot \bs v_{\bs k}^0 -  e (\bs B \cdot \bs \Omega_{n\bs k})  (\bs E \cdot \bs v_{\bs k}^0 ) \nonumber \\
    & - (\bs E \cdot \bs \nabla_{\bs k} ( \bs B \cdot \bs m_{n \bs k})) + e (\bs E \cdot \bs B)(\bs \Omega_{\bs k} \cdot \bs v_{n \bs k}^{(0)} )\nonumber \\  
    &+e\tau (\bs v_{n\bs k}^{(0)} \times \bs B) \cdot \bs \nabla_{\bs k} (\bs E \cdot \bs v_{n \bs k}^{0}) \Big] \partial_{\varepsilon} \tilde f_n^{(0)} \nonumber \\
    & - \hbar (\bs E \cdot \bs v_{n\bs k}^{0} )(\bs B\cdot \bs m_{n\bs k}) \partial^2_{\varepsilon}\tilde f_{n}^{(0)}  \Bigg], \label{distribution}
\end{align}
where $\hbar \bs v_{n\bs k}^{0} = \bs \nabla_{\bs k} \varepsilon^0_{n \bs k}$, and the $0$'th order distribution-functions $\tilde f_{n}^{(0)}$ only depend on the field-independent electronic dispersions $\varepsilon^0_{n \bs k}$.

Perturbatively expanding $\bs j_n$ now in powers of $\bs B$
\begin{align}
    \bs j_n = \sum_{i=0} \bs j^{(i)}_n
\end{align}
with $\bs j^{(i)}_n \propto B^i$ and plugin Eqs.~\eqref{EoM} and \eqref{distribution} into Eq.~\eqref{Currentdensity} we obtain:
\begin{align}
    \bs j^{(0)} = -\frac{e^2}{\hbar}  &\int_{n\bs k} \Big[ \tilde f^{(0)}_n \bs E \times \bs \Omega_{n\bs k}\nonumber \\
    &+ \hbar \tau \partial_\varepsilon \tilde f^{(0)}_n (\bs E \cdot \bs v_{n\bs k}^0)  \bs v_{n\bs k}^0 \Bigg], \label{j0}
\end{align}
that corresponds to the AHE response and Drude conductivity as first and second terms, respectively, and 
\begin{align}
    \bs j_n^{(1)} &=  \frac{e^2}{\hbar} \int_{n\bs k} \Big[ (\bs B \cdot \bs m_{n \bs k} ) (\bs E \times \bs \Omega_{n \bs k}) \partial_\varepsilon\tilde f^{(0)}_n \Big]  \nonumber \\
    &+ \frac{e^2 \tau}{\hbar}\int_{n\bs k} \partial_\varepsilon \tilde f^{(0)}_n \Bigg[ (\bs E \cdot \bs v_{n \bs k}^0 )\bs \nabla_{\bs k} (\bs B \cdot \bs m_{n \bs k} )   \nonumber \\
    &+ \bs v_{n \bs k}^0  (\bs E \cdot \bs \nabla_{\bs k}(\bs B \cdot \bs m_{n \bs k})) + e \big[ \bs v_{n \bs k}^0  (\bs B \cdot \bs \Omega_{n\bs k}) (\bs E \cdot \bs v_{n \bs k}^0 )  \nonumber \\
    &- \bs v_{n \bs k}^0  (\bs E \cdot \bs B) (\bs \Omega_{n \bs k} \cdot \bs v_{n \bs k}^0 )  - \bs B (\bs E \cdot \bs v_{n \bs k}^0 )(\bs \Omega_{n \bs k} \cdot \bs v_{n \bs k}^0 )  \big] \Bigg] \nonumber \\
    & + e^2\tau \int_{n\bs k}  \bs v_{n \bs k}^0  \Bigg[\partial^2_\varepsilon \tilde f^{(0)}_n (\bs E \cdot \bs v_{n \bs k}^0 )(\bs B \cdot \bs m_{n \bs k})  \nonumber \\
    &- \frac{e\tau}{\hbar} \partial_\varepsilon \tilde f^{(0)}_n(\bs v_{n \bs k}^0  \times \bs B) \cdot \bs \nabla_{\bs k} (\bs E \cdot \bs v_{n \bs k}^0 )
\Bigg]. \label{j1full}
    \end{align}

\eqref{j0} which expressed in terms of its associated symmetric and antisymmetric conductivity tensors yields
\begin{align}
    \sigma^{S}_{ij} = - e^2  \tau \int_{n,\bs k} \partial_\varepsilon\tilde f^0_n v^i_{n} v^j_{n}, \label{DRUDE}  
\end{align}
and
\begin{align}
    \sigma^{A}_{ij} =  - \frac{e^2}{\hbar}\epsilon_{ijk} \int_{n, \bs k} \tilde f_n^0\Omega_k , \label{HALL}  
\end{align}
as the the Drude and anomalous conductivity tensors, respectively, whereas the symmetrization in the $i$ and $j$ cartesian indices of Eq~\eqref{j1full} leads to
\begin{align}
\sigma^{S}_{ijk} =&\frac{\tau e^2}{\hbar}  \int_{n,k}  \partial_\varepsilon\tilde f^0_n \Big[  \frac{v^i_{n} \partial_jm^k_{n} + v^j_{n} \partial_i m^k_{n}}{2} \nonumber\\
&- v^{ij}_{n} m^k_{n} + e \Big( v^i_{n}  v^j_{n}  \Omega^k_{n} \nonumber\\ 
&- \left( \delta_{jk} v^i_{n} + \delta_{ik} v^j_{n}  \right)\sum_{q \in{x,y}}  v^q_{n} \Omega^q_{n}\Big) \Big], \label{semiclassical}
\end{align}
% \begin{align}
%     \sigma_{ij} = \sigma^{(0)}_{ij} + \sigma^{(1)}_{ijk} B_k  + \mathcal O (B^2),
% \end{align}
the general expression for the time-odd linear magnetoresponse studied in this work\cite{Morimoto16, Mandal22, Sunko:PRB25}, with $v_{ij,n} = \partial_i v_{j,n}$. Note that for the sake of simplicity the momentum dependency has been omitted in the previous expressions.

\subsection{LMC in the Canonical Ensemble}

Eq. \eqref{semiclassical} implicitly assumes the grand canonical ensemble, where the chemical potential is fixed and unaffected by $\bs B$. In this ensemble, the electron density is thus a function of $B$. Another possibility is to fix $n$ instead, in such a way the chemical potential is now $B$-dependent. The analogous expression to Eq. \eqref{semiclassical} in the canonical ensemble must then include a correction which is given by\cite{Sunko:PRB25}:
\begin{align}
    \delta j_1 =& - \frac{e^2}{\hbar} \delta \mu \int_{n\bs k}  \partial_\varepsilon \tilde f_n^{(0)} \bs E_\omega \times \bs \Omega_{n\bs k} \nonumber \\&+ \hbar \tau \partial^2_\varepsilon \tilde f_n^{(0)} (\bs E \cdot \bs v_{n\bs k}^0) \bs v_{n\bs k}^0,
    \label{jcanonicalcorrection}
\end{align}
with 
\begin{align} \delta \mu = \frac{1}{N_0} \int_{n\bs k}  \big[ \frac{e}{\hbar} (\bs B \cdot \bs \Omega_{n\bs k}) -(\bs B \cdot \bs m_{n\bs k}) \partial_\varepsilon  \big]  \tilde{f}_n^{(0)},
\end{align}
the $B$-induced chemical potential shift ($\mathcal O(B^1)$) and $N_0 =-\int_{n\bs k} \partial_\varepsilon \tilde f^{(0)}_n$. The first term in Eq. \eqref{jcanonicalcorrection} vanishes upon symmetrization due to the cross product and therefore the correction to Eq.~\eqref{semiclassical} reduces to
\begin{align}
\delta \sigma^S_{ijk} = \frac{-e^2 \tau}{N_0}  \int_{\bs k} \big[\frac{e}{\hbar}\Omega^k_{n}   - m^k_{n} \partial_\varepsilon \big] \tilde f^0_n\int_{\bs k}\partial^2_\varepsilon \tilde f^0_n v^i_{n} v^j_{n}. \label{correction}
\end{align}

\section{Evaluation of the semiclassical equations. The length gauge formalism for quasi two-dimensional systems\label{lengthgaugeexpressions}}

Semiclassical expressions can be derived from the density matrix formalism in the length gauge\cite{Aversa95} in the limit where interband coherence is neglected. In this framework, the electric field couples the position operator as $-e \bs E \cdot \bs r$, which drives carrier motion within a given band by changing the crystal momentum. This field-induced intraband dynamics can be described equivalently in two complementary ways: microscopically, through the semiclassical evolution of the wave-packet center ($\bs r$, $\bs k$) and statistically, through the drift term in the Boltzmann kinetic equation for the carrier distribution. This formulation avoids the gauge inconsistencies  associated with band truncation or origin dependence that may arise in alternative velocity gauge representations, and is therefore the one chosen in our calculations.

We briefly introduce the general length-gauge formalism in the following (we refer the interested reader to Refs.~\cite{Aversa95} for more details) and use it to derive the sum rules employed in the evaluation of Eq.~\eqref{fullsemiclassical} for the quasi-two-dimensional systems studied in this paper: TBG and ABC-stacked N-layer graphene.

\subsection{The length-gauge formalism}

The position operator $\boldsymbol r$ representation in the basis of Bloch states can be decomposed into intraband and an interband terms as follows:
\begin{align}
\boldsymbol r = \boldsymbol r_{\text{intra}} + \boldsymbol r_{\text{inter}},
\end{align}
where
\begin{align}
&\langle n \boldsymbol k| \boldsymbol r_{\text{intra}} | m \boldsymbol k' \rangle = \delta_{nm} \left[ \delta(\boldsymbol k - \boldsymbol k')A_{nn} + i \boldsymbol \nabla_{\boldsymbol k} \delta(\boldsymbol k - \boldsymbol k') \right], \nonumber\label{rintra} \\
&\langle n \boldsymbol k| \boldsymbol r_{\text{inter}} | m \boldsymbol k' \rangle = (1-\delta_{nm})  \delta(\boldsymbol k - \boldsymbol k')A_{nm}, 
\end{align}
and 
\begin{align}
    A_{ nm}(\boldsymbol k) = \frac{(2\pi)^3 i}{V} \int_V [dr] u_{n\boldsymbol k}^*(\boldsymbol r) \boldsymbol \nabla_{\boldsymbol k}u_{m\boldsymbol k}(\boldsymbol r),
\end{align} denotes the Berry connection of bands $n$ and $m$. It is important to note that the intraband Bloch representation of the position operator is ill-defined due to: (1) the non-uniqueness of $A_{nn}$ and (2) the singular distributional character of its derivative term which stems from the extended nature of the Bloch states. These two pose an apparent challenge for the evaluation of observables in the length gauge, provided the explicit dependence of the perturbed Hamiltonian on $\bs r_{nm}$ through the dipolar coupling. However, the so-called Blount's Theorem\cite{Blount} offers an elegant solution.
Let $S$ be an operator diagonal in $k$, then
\begin{align} &\langle n, \boldsymbol k| [r^a_{\text{intra}}, S] |m, \boldsymbol k' \rangle = i S_{nm;a} \delta(\boldsymbol k- \boldsymbol k'), \label{Blountstheorem1}
\end{align}
with 
\begin{align}
S_{nm;a}= \partial_{k_a}S_{nm} - i  S_{nm} (A_{nn}-A_{mm}).
\label{Blountstheorem2}
\end{align}
This relationship is extremely useful for those expressions where the dependency on $\bs r_{\text{intra}}$ manifests inside a commutator with a well-defined (diagonal in $k$) operator $S$. Also crucially, $S_{nm;a}$ is uniquely-defined even though the individual terms in Eq. \eqref{Blountstheorem2} are not. This sometimes-called Blount's prescription will be employed in the following sections to derive sum rules for the integrand of Eq.~\eqref{fullsemiclassical}.

Importantly, this framework assumes unbounded directions in three dimensions, in the following, we will derive general expressions based on the molecular limit\cite{Pozo23} for quasi two-dimensional systems. Specifically, the following sections will deal with $\Omega_{n,i}$, $m_{n, i}$, and $\partial_{k_i} m_{n,i}$ where $i\in \{x,\ y\}$ and $z$ is the bounded direction, since they are the relevant quantities for the in-plane LMC.

\subsection{In-plane Berry curvature}

Here we derive, using the Blount's prescription, the expression for the in-plane Berry curvature of a two-dimensional layered system bounded in the $z$ direction. In the following, we adopt the Einstein summation, with greek and latin indices $\in \{x,y,z\}$ and $\in \{x,y\}$, respectively. The bulk expression reads:
\begin{align}
    \Omega^i = \epsilon^{ij} \partial_j  \langle n | r^z | n \rangle,
    \label{generalBC}
\end{align}
with $\partial_i = \partial_{k_i}$. Using the position operator decomposition and Blount's theorem (in Eq.~\eqref{Blountstheorem1}), the k-derivative dependence can be eliminated as follows:
\begin{align}
[r^\alpha_{\text intra}, r^\beta_{\text intra}] &= [r^\alpha -r^\alpha_{\text inter} , r^\beta - r^\beta_{\text inter}] \nonumber\\ 
&= [r^\alpha, r^\beta]+ [ r^\alpha_{\text inter} , r^\beta_{\text inter}]\nonumber \\ &- ( [ r^\alpha , r^\beta_{\text inter}]+ [ r^\alpha_{\text inter} , r^\beta]) 
\end{align}
with 
\begin{align}
[r^\alpha &, r^\beta_{\text inter}]_{nm}+ [ r^\alpha_{\text inter} , r^\beta]_{nm}  
	 = \nonumber \\ 
     &\sum_q r^\alpha_{nq}r^\beta_{qm}(1-\delta_{qm}) 
     - r^\beta_{nq}r^\alpha_{qm}(1-\delta_{nq}) \nonumber \\
     &+r^\beta_{nq}r^\alpha_{qm}(1-\delta_{qm})-r^\alpha_{nq}r^\beta_{qm}(1-\delta_{qn}), 
\end{align}
which vanish for diagonal matrix elements, therefore
 \begin{align}
     [r^\alpha_{\text intra}, r^\beta_{\text intra}]_{nn} &= [ r^\alpha_{\text inter} , r^\beta_{\text inter}]_{nn}\nonumber \\ &= \sum_{q\neq n} r^\alpha_{nq} r^\beta_{qn} - r^\beta_{nq} r^\alpha_{qn}. 
 \end{align}
By setting $\beta = z$ we can apply the Blount's theorem in Eq. \eqref{Blountstheorem1}, provided $r^z$ is non singular when $\bs k = \bs k'$, to obtain:
\begin{align}
    [r^\alpha_{\text intra}, r^z_{\text intra}]_{nn} = i r_{nn;\alpha}^z= i \partial_\alpha r_{nn}^z. \label{partialalpharz}
\end{align}

Substituting Eq. \eqref{partialalpharz} into Eq. \eqref{generalBC}, we obtain
\begin{align}
    \Omega^i_{nn} = 2\epsilon^{i j}\text{Im}\bigg[\sum_{n'\neq n}r^j_{nq}r^z_{qn}\bigg],
\end{align}
which coincides with the expression found in Refs. \cite{Ghorai25, Kim21}.

\subsection{Planar magnetic moment and its gradient}
The bulk expression for the $i$'th component of the magnetic orbital moment operator of the band $n$ in ``origin-independent" form~\cite{Pozo23} is given by:
\begin{align}
 \langle n | m^i |n \rangle \equiv m^i_{nn} &= \frac{-e}{2\hbar} 
 \epsilon^{ijk}  (r^k_{nm} - r^k_{nn}\delta_{nm} )v^j_{mn} 
\end{align}

Taking the molecular limit, we can write the corresponding expression for a planar two-dimensional system bounded in $z$ as follows:
\begin{align}
    m^i_{nn} &= \frac{-e}{2\hbar} \epsilon^{ij} \Bigg[ \sum_q (r^j_{nq}-r^j_{nn}\delta_{nq}) v_{qn}^z \nonumber \\
    & \ \ \ \ \ \ \ \ \ \ \ \ \ \ \  - \sum_q (r^z_{nq} - r^z_{nn}\delta_{nq})v_{qn}^j \Bigg] \nonumber \\
    &=  \frac{-e}{2\hbar}\epsilon^{ij} \sum_{q\neq n}\left[  r^j_{nq} v_{qn}^z - r^z_{nq}v_{qn}^j \right], \label{generalOMM}
\end{align}
where $\langle n | v^z | n \rangle = 0$ , although $\langle n | v^z | m \rangle \neq 0$ in general. Then: 
\begin{align}
  &\sum_{q\neq n}\Big[  r^j_{nq} v_{qn}^z - r^z_{nq}v_{qn}^j \Big] =    \sum_{q\neq n}\left[-ir^j_{nq} [r^z, H_0]_{qn}/\hbar - r^z_{nq} v^j_{qn}\right] \nonumber \\
  &=  \sum_{q\neq n} \left[-i r_{nq}^j [r^z_{\text intra} + r^z_{\text{inter}}, H_0]_{qn}/\hbar- r^z_{nq} v^j_{qn}\right],
  \label{intermOMM}
\end{align}     
where we have used that $\dot{r}^\alpha = -i[r^\alpha, H]/\hbar = -i[r^\alpha, H_0]/\hbar +ieE[r^\alpha, r^\beta]/\hbar = -i[r^\alpha, H_0]/\hbar$,

since $[r^\alpha, r^\beta]=0$ and have decomposed the k-space representation of the position operators in terms of intraband and interband matrix elements. We now get rid of the singular  intraband representations of $r$ using sum rules:

Since $\{|n\rangle \}$ constitutes a complete and orthonormal basis for $H_0$:
\begin{align}
[r^\alpha_{\text intra},H_0]_{nm} =  
    \begin{cases} 
i\partial_\alpha H_{0,nm}\delta_{nm}, & \text{if } \alpha \in \{x,y\} \\ 
    0, & \alpha = z
\end{cases}
\end{align}
and
\begin{align}
    [r^\alpha_{\text{inter}}, H_0]_{nm} &= \sum_{p\neq n} r^\alpha_{np} H_{0,pm}-H_{0,np}r^\alpha_{pm} \nonumber \\ 
    &=  r^\alpha_{nm} (\epsilon_m -\epsilon_n).
\end{align}
The RHS of Eq. \eqref{intermOMM} reads
\begin{align}
  \sum_{q\neq n}&-ir_{nq}^j r^z_{qn} (\epsilon_n -\epsilon_q)/\hbar -\sum_{q\neq n} r^z_{nq}v^j_{qn}\nonumber \\
  &= -\sum_{q\neq n}\left[v_{nq}^j r^z_{qn} + r^z_{nq}  v^j_{qn}\right] \label{lastOMM}
\end{align}    
using that $r^j_{nm} = -i \hbar v^j_{nm}/ \epsilon_{nm}$ with $n\neq m$.  Finally, we substitute Eq. \eqref{lastOMM} into Eq. \eqref{generalOMM} to obtain:
\begin{align}
m^i_{nn} =\frac{e}{2\hbar} \epsilon^{ij} \left[\sum_{q\neq n} v^j_{nq} r_{qn}^z +  r^z_{nq} v^j_{qn}\right] = \frac{e}{\hbar}\epsilon^{ij}\text{Re} \sum_{q\neq n} v^j_{nq} r_{qn}^z \label{OMMlg}\end{align}
\subsection{$k$-derivatives of the orbital magnetic moment}

The semiclassical equation in Eq. \eqref{semiclassical} contains $k$-derivatives of the orbital magnetic moment. We employ the Blount's prescription \cite{Blount, Aversa95} once more to express them as a sum rule in terms of gauge invariant quantities:
\begin{align}\hbar\partial_i [v_{nq}^j r^z_{qn}]&=\partial_i [i \epsilon_{nm} r_{nm}^j r^z_{qn}]\nonumber \\
&=i\partial_i [ \epsilon_{nq}] r_{nq}^j r^z_{qn}+ i \epsilon_{nq}\partial_i [ r_{nq}^j r^z_{qn}] \nonumber \\ 
&= i \Delta^i_{nq}r_{nq}^j r^z_{qn} + i \epsilon_{nq}\partial_i [ r_{nq}^j r^z_{qn}], \label{chainrule}\end{align}
with $n\neq q$ and $\Delta^i_{nq} = \hbar(v^i_n-v^i_q)$. We now use Eq.~\eqref{Blountstheorem1} to obtain: \begin{align}r^j_{nq;i}=\partial_{i}r^j_{nq} - i r^j_{nq}(r^i_{nn}-r^i_{qq}). \label{covariant}\end{align}
It is important to note that only the covariant derivative in the previous expression is gauge invariant although each term in the RHS of Eq.~ \eqref{covariant} is not individually. Applying the chain rule to the last term of Eq. \eqref{chainrule}:
\begin{align}\partial_i [ r_{nq}^j r^\alpha_{qn}] &= \partial_i [ r_{nq}^j] r^\alpha_{qn}+  r_{nq}^j \partial_i [r^\alpha_{qn}] \nonumber \\
&= r^j_{nq;i}r^\alpha_{qn}+ i r^j_{nq}(r^i_{nn}-r^i_{qq})r^\alpha_{qn}\nonumber \\ &+r^j_{nq} r^\alpha_{qn;i}- i r^j_{nq} r^\alpha_{qn}(r^i_{nn}-r^i_{qq}) \nonumber \\
&= r^j_{nq} r^\alpha_{qn;i}+r^j_{nq;i}r^\alpha_{qn}.
\label{derivativeidentity}
\end{align}

Plugging Eqs. \eqref{chainrule} and \eqref{derivativeidentity} into Eq. \eqref{OMM} for $\alpha = z$ we obtain
\begin{align}
    \partial_im_{j,nn}&= -\frac{e}{\hbar^2}\epsilon^{jk}\text{Im} \Big[ \sum_{q\neq n} \Delta^j_{nq}r_{nq}^k r^z_{qn} \nonumber \\
    &+ \epsilon_{nq}[r^k_{nq} r^z_{qn;j}+r^k_{nq;j  }r^z_{qn}] \Big]  
\end{align}
where
\begin{align}
r^i_{nm;j}(k) &= -\frac{\hbar}{i\epsilon_{nm}} \Big[ \frac{v_{nm}^i \Delta_{mn}^j + v_{nm}^j\Delta_{mn}^i}{\epsilon_{nm}} - w^{ij}_{nm}\nonumber \\
&+ \hbar \sum_{p \neq,n,m} \frac{v^i_{np}v^j_{pm}}{\epsilon_{pm}}-\frac{v^j_{np}v^i_{pm}}{\epsilon_{np}} \Big],
\end{align} and
\begin{align}
r^z_{nm;j} &= \hbar \sum_{p\neq m} r^z_{np}\frac{v^j_{pm}}{\epsilon_{pm}} -  \hbar \sum_{p\neq n} \frac{v^j_{np}}{\epsilon_{np}}r^z_{pm},
\end{align}
are the generalized (covariant) derivatives along an in-plane direction of the in-plane and out-of-plane non-diagonal components of the position operator, respectively (details on the out-of-plane covariant derivative in Ref.~\cite{Penaranda24}).

\section{$r^z$ in TBG}\label{App:RzinTBG}

The representation of the $z$-component of the  position operator $r^z$ is required for the evaluation of the length gauge formulas derived in the previous section and, therefore, to compute the in-plane LMC response. Although trivial for N-layer RG, its definition in the THFM of TBGis not straightforward. The reason is that, as a result of the projection into the $f-c$ degrees of freedom, the layer index is lost. In this basis, it can be approximated by $r^z$\cite{Penaranda:PRB26}
\begin{align}
  r^z \approx 2 \xi \eta \sum_{i\in{0,1,2}} \sin(\bs k \cdot \bs a_n) f^\dagger_{\bs k, \alpha s \eta} f_{\bs k, \alpha s \eta}
\end{align}
with $f^\dagger_{\bs k, \alpha s \eta} = \frac{1}{\sqrt{N}} \sum_{\bs R} \exp(i \bs k \cdot \bs R) f^\dagger_{\bs R, \alpha s \eta}$. It corresponds to a non-local operator that preserves the symmetries of $r^z$, namely, it is time-even, particle-hole-odd, and transforms as an $A_2$ irreducible representation under TBG's point group (see Ref.~\cite{Penaranda:PRB26} for more details).

\begin{figure*}[t]
    \centering
    \includegraphics[width=1\linewidth]{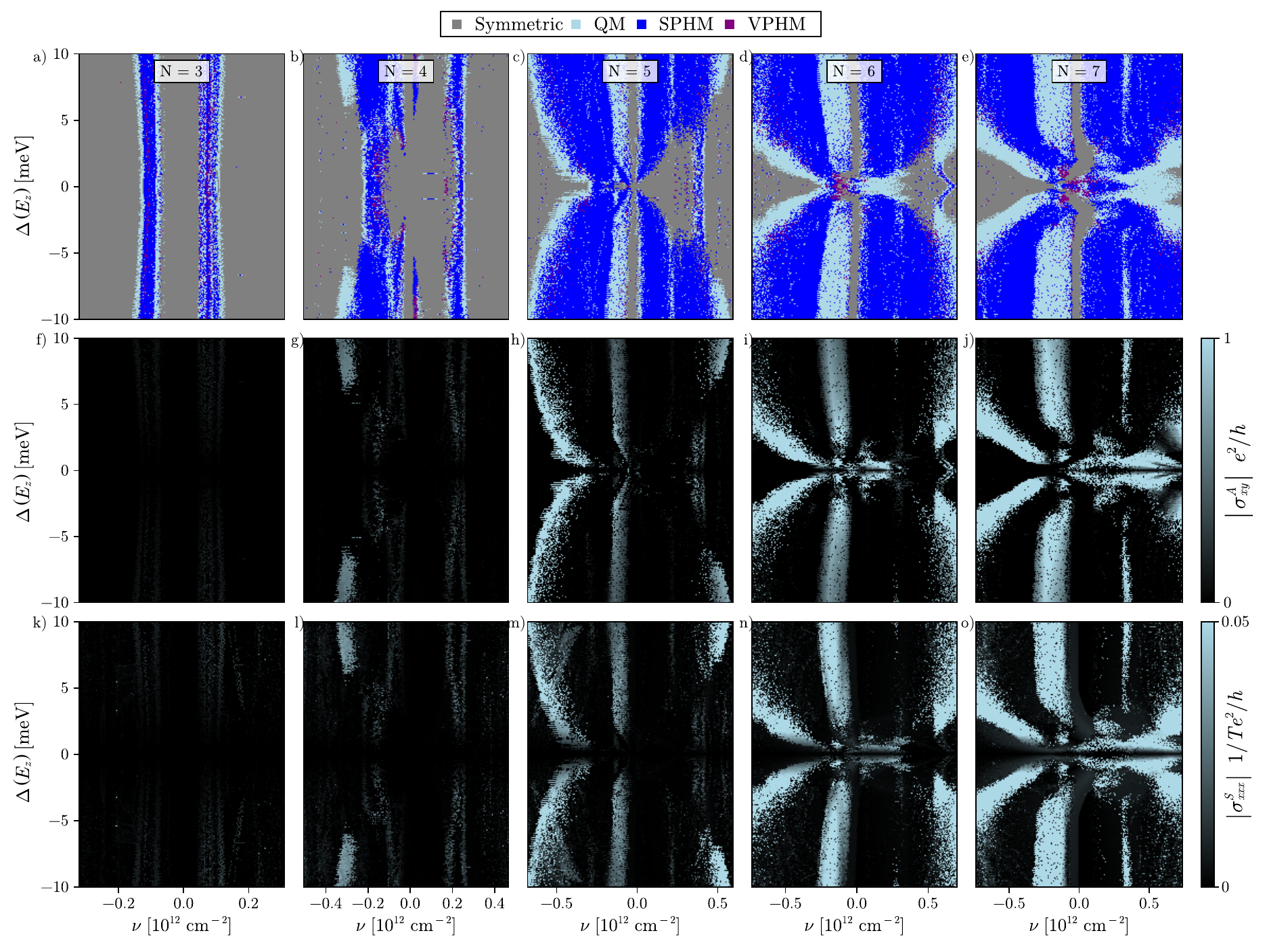}
    \caption{Transport probes of valley polarization in N-layer RG: LMC vs QAH. (a-e) Interacting phase diagrams $(\Delta_{E_z}$-$\nu)$ as a function of the number of layers $N = 3,4,5,6,7$, respectively. Symmetric (gray) refers to an $SU4$ preserving phase, and the $SU(2)$ SSB phases are QM (light blue), VPHM (blue), and SPHM (purple). Absolute values of the QAH (f-j) and LMC (k-o) responses as a function of layer index corresponding for the ground states in (a-e). Rest of parameters coincide with Fig.~\ref{fig:pdmain}.}
    \label{fig:PDs}
\end{figure*}

\section{LMC vs QAH as a function of layer index in rhombohedral graphene multilayers}\label{NlayerRGLMC}
A comparison between the LMC and QAH in N-layer RG is presented in Fig.~\ref{fig:pdmain} as function of the layer number, with $N=3,4,5,6,7$. Panels (a-e) show the interacting phase diagram ($\nu$ vs $\Delta(E_z)$) featuring phase transitions between symmetric, QM, SPHM, and VPHM phases. In (f-j) and (k-o) we show the AHE and LMC in absolute values associated with (a-e). We find, irrespective of $N$, similar results as those described in the main text for $N=7$ (last column), namely, the LMC is only finite in VP phases and the LMC response at moderate fields is of the same order of magnitude than the AHE, where the different functional dependencies across the phase diagram (see \ref{LMCnonintRG}) may switch the dominance of one probe of broken TRS with respect to the other.
\bibliography{transport}
\end{document}